%% file: HIG-18-005_temp.tex
\pdfoutput=1

\documentclass[11pt,twoside,a4paper,cmspaper,final,collab]{cms-tdr}
\def\svnVersion{ac6ccf5-D}\def\svnDate{2019/07/05}\def\cmsCernNoTag{CERN-EP-2018-343}\def\cmsCernDate{\today}\def\cmsMessage{"Published in the European Physical Journal C as \href{http://dx.doi.org/10.1140/epjc/s10052-019-7058-z}{\doi{10.1140/epjc/s10052-019-7058-z}.}"}

\begin{document}\cmsNoteHeader{HIG-18-005}

\hyphenation{had-ron-i-za-tion}
\hyphenation{cal-or-i-me-ter}
\hyphenation{de-vices}
\RCS$HeadURL$
\RCS$Id$

\newlength\cmsFigWidth
\ifthenelse{\boolean{cms@external}}{\setlength\cmsFigWidth{0.49\textwidth}}{\setlength\cmsFigWidth{0.65\textwidth}}
\ifthenelse{\boolean{cms@external}}{\providecommand{\cmsLeft}{upper\xspace}}{\providecommand{\cmsLeft}{left\xspace}}
\ifthenelse{\boolean{cms@external}}{\providecommand{\cmsRight}{lower\xspace}}{\providecommand{\cmsRight}{right\xspace}}
\providecommand{\cmsTable}[1]{\resizebox{\textwidth}{!}{#1}}
\newlength\cmsTabSkip\setlength{\cmsTabSkip}{1ex}

\newcommand{\V}{\ensuremath{\cmsSymbolFace{V}}\xspace}
\newcommand{\A}{\ensuremath{\cmsSymbolFace{A}}\xspace}
\newcommand{\VV}{\ensuremath{\V\V}\xspace}
\newcommand{\ST}{\ensuremath{\PQt\text{+X}}\xspace}
\newcommand{\Vjets}{\ensuremath{\V\text{+jets}}\xspace}
\newcommand{\Wjets}{\ensuremath{\PW\text{+jets}}\xspace}
\newcommand{\Zjets}{\ensuremath{\PZ\text{+jets}}\xspace}
\newcommand{\mZH}{\ensuremath{m_{\PZ\Ph}}\xspace}
\newcommand{\mtZH}{\ensuremath{m_{\PZ\Ph}^{\text{T}}}\xspace}
\newcommand{\mA}{\ensuremath{m_{\A}}\xspace}
\newcommand{\mH}{\ensuremath{m_{\PH}}\xspace}
\newcommand{\mHpm}{\ensuremath{m_{\PH^\pm}}\xspace}
\newcommand{\mh}{\ensuremath{m_{\Ph}}\xspace}
\newcommand{\mjj}{\ensuremath{m_\mathrm{jj}}\xspace}
\newcommand{\mbb}{\ensuremath{m_{{\cPqb}{\cPqb}}}\xspace}
\newcommand{\mll}{\ensuremath{m_{\ell\ell}}\xspace}
\newcommand{\B}{\ensuremath{\mathcal{B}}}
\newcommand{\cosba}{\ensuremath{\cos(\beta-\alpha)}\xspace}

\cmsNoteHeader{HIG-18-005}
\title{Search for a heavy pseudoscalar boson decaying to a Z and a Higgs boson at $\sqrt{s}=13\TeV$}

\date{\today}

\abstract{
   A search is presented for a heavy pseudoscalar boson \A decaying to a \PZ boson and a Higgs boson with mass of 125\GeV. In the final state considered, the Higgs boson decays to a bottom quark and antiquark, and the \PZ boson decays either into a pair of electrons, muons, or neutrinos. The analysis is performed using a data sample corresponding to an integrated luminosity of 35.9\fbinv collected in 2016 by the CMS experiment at the LHC from proton-proton collisions at a center-of-mass energy of 13\TeV. The data are found to be consistent with the background expectations. Exclusion limits are set in the context of two-Higgs-doublet models in the \A boson mass range between 225 and 1000\GeV.
}

\hypersetup{%
pdfauthor={CMS Collaboration},%
pdftitle={Search for a heavy pseudoscalar boson decaying to a Z boson and a Higgs boson at sqrt(s)=13 TeV},%
pdfsubject={CMS},%
pdfkeywords={CMS, physics, software, computing}}

\maketitle

\section{Introduction}

The discovery of a Higgs boson at the CERN LHC~\cite{bib:Aad20121,bib:Chatrchyan201230} and the measurement of its mass, spin, parity, and couplings~\cite{Aad:2015zhl,Khachatryan:2016vau} raises the question of whether the Higgs boson sector consists of only one scalar doublet, which results in a single physical Higgs boson as expected in the standard model (SM), or whether additional bosons are involved in electroweak (EW) symmetry breaking.

The two-Higgs-doublet model (2HDM)~\cite{Branco:2011iw} provides an extension of the SM Higgs boson sector introducing a second scalar doublet. The 2HDM is incorporated in supersymmetric models~\cite{Martin:1997ns}, axion models~\cite{Kim:1986ax}, and may introduce additional sources of explicit or spontaneous \textit{CP} violation that explain the baryon asymmetry of the universe~\cite{Fromme:2006cm}.
Various formulations of the 2HDM predict different couplings of the two doublets to right-handed quarks and charged leptons:
in the Type-I formulation, all fermions couple to only one Higgs doublet; in the Type-II formulation, the up-type quarks couple to a different doublet than the down-type quarks and leptons; in the ``lepton-specific'' formulation, the quarks couple to one of the Higgs doublets and the leptons couple to the other; and in the ``flipped'' formulation, the up-type fermions and leptons couple to one of the Higgs doublets, while the down-type quarks couple to the other.

\begin{figure}[!htb]\centering
  \includegraphics[width=0.36\textwidth]{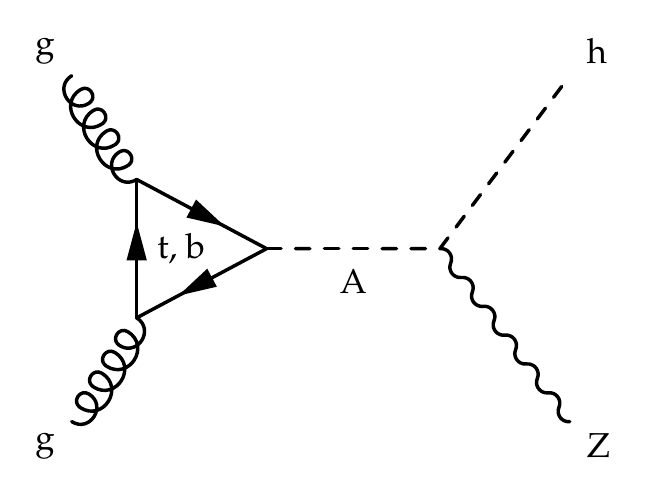}
  \includegraphics[width=0.36\textwidth]{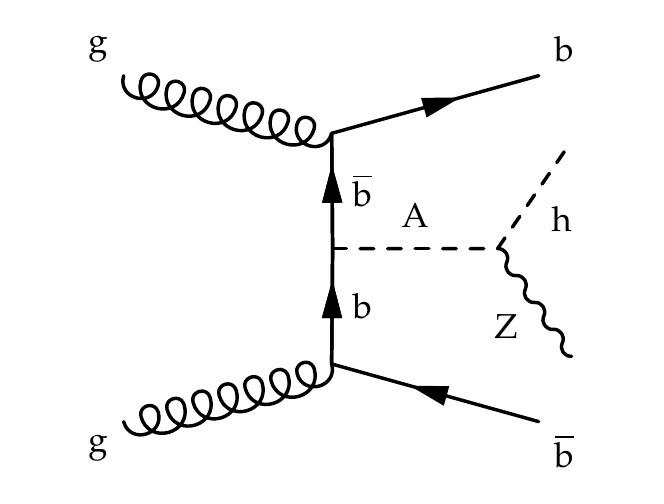}
  \caption{Representative Feynman diagrams of the production in the 2HDM of a pseudoscalar \A boson via gluon-gluon fusion (\cmsLeft) and in association with {\cPqb} quarks (\cmsRight).}
  \label{fig:Feynman}
\end{figure}

The two Higgs doublets entail the presence of five physical states: two neutral and \textit{CP}-even bosons (\Ph and \PH, the latter being more massive), a neutral and \textit{CP}-odd boson (\A), and two charged scalar bosons ($\PH^\pm$). 
The model has two free parameters, $\alpha$ and $\tanb$, which are the mixing angle and the ratio of the vacuum expectation values of the two Higgs doublets, respectively. If $\tanb \lesssim 5$, the dominant \A boson production process is via gluon-gluon fusion, otherwise associated production with a {\cPqb} quark-antiquark pair becomes significant. The diagrams of the two production modes are shown in Fig.~\ref{fig:Feynman}. At small \tanb values the heavy pseudoscalar boson \A may decay with a large branching fraction to a \PZ and an \Ph boson, if kinematically allowed~\cite{Branco:2011iw}.
These models can be probed either with indirect searches, by measuring the cross section and couplings of the SM Higgs boson~\cite{Sirunyan:2018koj}, or by performing a direct search for an \A boson.

This paper describes a search for a heavy pseudoscalar \A boson that decays to a \PZ and an \Ph boson, both on-shell, with the \PZ boson decaying to $\ell^+\ell^-$ ($\ell$ being an electron or a muon) or to a pair of neutrinos, and the \Ph boson to \bbbar. The \Ph boson is assumed to be the 125\GeV boson discovered at the LHC.
In this search, the candidate \A boson is reconstructed from the invariant mass of the visible decay products in events when the \PZ boson decays to charged leptons, or is inferred through a partial reconstruction of the mass using quantities measured in the transverse plane when the \PZ boson decays to neutrinos. The signal would emerge as a peak above the SM continuum of the four-body invariant mass (\mZH) spectrum for the former decay mode and the transverse mass (\mtZH) for the latter.
The signal sensitivity is maximized by exploiting the known value of the \Ph boson mass to rescale the jet momenta and significantly improve the \mZH resolution. In addition, selections based on multivariate discriminators, exploiting event variables such as angular distributions, are used to optimize the signal efficiency and background rejection.
This search is particularly sensitive to a pseudoscalar \A boson with a mass smaller than twice the top quark mass and for small \tanb values. In this region of the 2HDM parameter space, the \A boson cross section is larger than 1~\unit{pb}, and the \A boson decays predominantly to $\PZ\Ph$~\cite{Branco:2011iw}.

With respect to the CMS search performed at $\sqrt{s}=8\TeV$~\cite{Khachatryan:2015lba}, this analysis benefits from the increased center-of-mass energy and integrated luminosity, includes final states with invisible decays of the \PZ boson, increases the sensitivity to {\cPqb} quark associated production, and extends the \A boson mass (\mA) range from 600 to 1000\GeV. At larger \mA, the angular separation between the {\cPqb} quarks becomes small, and the Higgs boson is reconstructed as a single large-cone jet; the corresponding CMS analysis presents limits on the 2HDM from 800\GeV to 2\TeV~\cite{Sirunyan:2018qob}. The ATLAS Collaboration has published a search probing $\PZ\Ph$ resonances with similar event selections based on a comparable data set, observing a mild excess near 440\GeV in categories with additional {\cPqb} quarks~\cite{Aaboud:2017cxo}.

\section{The CMS detector}\label{sec:detector}

A detailed description of the CMS detector, together with a definition of the coordinate system used and the relevant kinematic variables, can be found in Ref.~\cite{Chatrchyan:2008zzk}.

The central feature of the CMS apparatus is a superconducting solenoid of 6\unit{m} internal diameter, providing a magnetic field of 3.8\unit{T}. Within the solenoid volume are a silicon pixel and strip tracker, a lead tungstate crystal electromagnetic calorimeter (ECAL), and a brass and scintillator hadron calorimeter (HCAL), each composed of a barrel and two endcap sections. Forward calorimeters extend the pseudorapidity coverage provided by the barrel and endcap detectors. Muons are detected in gas-ionization chambers embedded in the steel flux-return yoke outside the solenoid. 

The silicon tracker measures charged particles within the pseudorapidity range $\abs{\eta} < 2.5$. It consists of 1440 silicon pixel and 15\,148 silicon strip detector modules. For nonisolated particles with transverse momenta of $1 < \pt < 10\GeV$ and $\abs{\eta} < 1.4$, the track resolutions are typically 1.5\% in \pt and 25--90 (45--150)\mum in the transverse (longitudinal) impact parameter~\cite{Chatrchyan:2014fea}.
The ECAL provides coverage up to $\abs{\eta} < 3.0$, and the energy resolution for unconverted or late-converting electrons and photons in the barrel section is about 1\% for particles that have energies in the range of tens of \GeV. The dielectron mass resolution for $\PZ\to\Pep\Pem$ decays when both electrons are in the ECAL barrel is 1.9\%, and is 2.9\% when both electrons are in the endcaps~\cite{Khachatryan:2015hwa}.
The muon detectors covering the range $\abs{\eta}< 2.4$ make use of three different technologies: drift tubes, cathode strip chambers, and resistive-plate chambers. Combining muon tracks with matching tracks measured in the silicon tracker results in a \pt resolution of 2--10\% for muons with $0.1 < \pt < 1\TeV$~\cite{Chatrchyan:2012xi}.

The first level of the CMS trigger system~\cite{Khachatryan:2016bia}, composed of custom hardware processors, uses information from the calorimeters and muon detectors to select the most interesting events in a fixed time interval of less than 4\mus. The high-level trigger (HLT) processor farm decreases the event rate from around 100\unit{kHz} to about 1\unit{kHz}, before data storage.

\section{Event reconstruction}\label{sec:reco}
{\tolerance=8000
A global event reconstruction is performed with a particle-flow (PF) algorithm~\cite{Sirunyan:2017ulk}, which uses an optimized combination of information from the various elements of the  detector to identify stable particles reconstructed in the detector as an electron, a muon, a photon, a charged or a neutral hadron. The PF particles have to pass the charged-hadron subtraction (CHS) algorithm~\cite{CMS-PAS-JME-14-001}, which discards charged hadrons not originating from the primary vertex, depending on the longitudinal impact parameter of the track. The primary vertex is selected as the vertex with the largest value of summed $\pt^2$ of the PF particles, including charged leptons, neutral and charged hadrons clustered in jets, and the associated missing transverse momentum \ptvecmiss, which is the negative vector sum of the \ptvec of those jets.

Electrons are reconstructed in the fiducial region $\abs{\eta}<2.5$ by matching the energy deposits in the ECAL with charged particle trajectories reconstructed in the tracker~\cite{Khachatryan:2015hwa}. The electron identification is based on the distribution of energy deposited along the electron trajectory, the direction and momentum of the track, and its compatibility with the primary vertex of the event.
Electrons are further required to be isolated from other energy deposits in the detector.
The electron relative isolation parameter is defined as the sum of transverse momenta of all the PF candidates, excluding the electron itself, divided by the electron \pt. The PF candidates are considered if they lie within $\Delta R = \sqrt{\smash[b]{(\Delta\eta)^2+(\Delta\phi)^2}} < 0.3$ around the electron direction, where $\phi$ is the azimuthal angle in radians, and after the contributions from pileup and other reconstructed electrons are removed~\cite{Khachatryan:2015hwa}.

Muons are reconstructed within the acceptance of the CMS muon systems using tracks reconstructed in both the muon spectrometer and the silicon tracker~\cite{Chatrchyan:2012xi}. 
Additional requirements are based on the compatibility of the trajectory with the primary vertex, and on the number of hits observed in the tracker and muon systems.
Similarly to electrons, muons are required to be isolated.
The muon isolation is computed from reconstructed PF candidates within a cone of $\Delta R< 0.4$ around the muon direction, ignoring the candidate muon, and divided by the muon \pt~\cite{Chatrchyan:2012xi}.

Hadronically decaying $\tau$ leptons are used to reject $\PW\to\tau\nu$ background events, and are reconstructed by combining one or three hadronic charged PF candidates with up to two neutral pions, the latter also reconstructed by the PF algorithm from the photons arising from the $\Pgpz \to\gamma\gamma$ decay~\cite{Khachatryan:2015dfa}.

Jets are clustered using the anti-\kt algorithm~\cite{Cacciari:2008gp,Cacciari:2011ma} with a distance parameter of $0.4$.
The contribution of neutral particles originating from pileup interactions is estimated to be proportional to the jet area derived using the {\FASTJET} package~\cite{Cacciari:2011ma,Cacciari:2008gn}, and subtracted from the jet energy.
Jet energy corrections, extracted from both simulation and data in multijet, $\gamma$+jets, and $\PZ$+jets events, are applied as functions of the \pt and $\eta$ of the jet to correct the jet response and to account for residual differences between data and simulation. The jet energy resolution amounts typically to $15-20\%$ at 30\GeV, 10\% at 100\GeV, and 5\% at 1\TeV~\cite{Khachatryan:2016kdb}.

Jets that originate from {\cPqb} quarks are identified with a combined secondary vertex {\cPqb}-tagging algorithm~\cite{BTV-16-002} that uses the tracks and secondary vertices associated with the jets as inputs to a neural network. The algorithm provides a {\cPqb} jet tagging efficiency of 70\%, and a misidentification rate in a sample of quark and gluon jets of about 1\%. The {\cPqb} tagging efficiency is corrected to take into account a difference at the few percent level in algorithm performance for data and simulation~\cite{BTV-16-002}.\par}

\section{Data and simulated samples}\label{sec:mcsimulation}
{\tolerance=8000
The data sample analyzed in this search corresponds to an integrated luminosity of 35.9\fbinv of proton-proton ({\Pp}{\Pp})~collisions at a center-of-mass energy of 13\TeV collected with the CMS detector at the LHC. Data are collected using triggers that require either the presence of at least one isolated electron or isolated muon with $\pt>27\GeV$, or alternatively a \ptmiss or \mht larger than 90--110\GeV, the value depending on the instantaneous luminosity. The \ptmiss is the magnitude of \ptvecmiss, and \mht is defined as the momentum imbalance of the jets in the transverse plane~\cite{Khachatryan:2016bia}.

The pseudoscalar boson signal is simulated at leading order (LO) with the \MGvATNLO 2.2.2 matrix element generator~\cite{MadGraph} in both the gluon-gluon fusion and {\cPqb} quark associated production modes according to the 2HDM~\cite{Branco:2011iw}, assuming a narrow signal width. The \Ph boson mass is set to 125\GeV, and the \A boson mass ranges between 225 and 1000\GeV. The $\A\to\PZ\Ph$ decay is simulated with \textsc{MadSpin}~\cite{Artoisenet:2012st}.
The Higgs boson is forced to decay to \bbbar, and the vector boson to a pair of electrons, muons, $\tau$ leptons, or neutrinos. In the gluon-gluon fusion production mode, up to one additional jet is included in matrix element calculations, and only the top quark contributes to the loop shown in Fig.~\ref{fig:Feynman} (\cmsLeft). The 2HDM cross sections and branching fractions are computed at next-to-next-to-leading order (NNLO) with \textsc{2hdmc} 1.7.0~\cite{Eriksson2010189} and \textsc{SusHi} 1.6.1~\cite{Harlander20131605}, respectively. The parameters used in the models are: $\mh=125\GeV$, $\mH=m_{\PH^\pm}=\mA$, the discrete $\mathrm{Z}_2$ symmetry is broken as in the minimal supersymmetric standard model (MSSM), and \textit{CP} is conserved at tree level in the 2HDM Higgs sector~\cite{Branco:2011iw}. The branching fractions of the \PZ boson are taken from the measured values~\cite{PhysRevD.98.030001}.

The SM backgrounds in this search consist of the inclusive production of a vector boson in association with other jets (\Vjets, with $\V=\PW$ or $\PZ$, and \V decaying to final states with charged leptons and neutrinos), and top quark pair production (\ttbar). \Vjets events are simulated at LO with \MGvATNLO with up to four partons included in the matrix element calculations and using the \textsc{MLM} matching scheme~\cite{Alwall:2007fs}. The event yield is normalized to the NNLO cross section computed with {\FEWZ} v3.1~\cite{Li:2012wna}. The \V boson \pt spectra are corrected to account for next-to-leading order (NLO) quantum chromodynamics (QCD) and EW contributions~\cite{Kallweit:2015dum}. The \ttbar and single top quark in the $t$ channel and $\PQt\PW$ production are simulated at NLO with \POWHEG v2 generator~\cite{Nason:2004rx,Frixione:2007vw,Alioli:2010xd}. The number of events for the top quark pair production process is rescaled according to the cross section computed with \textsc{Top++} v2.0~\cite{Czakon:2011xx} at NNLO+NNLL, and the transverse momenta of top quarks are corrected to match the distribution observed in data~\cite{Khachatryan:2016mnb}. Other SM processes, such as SM vector boson pair production (\VV), SM Higgs boson production in association with a vector boson ($\V\Ph$), single top quark (\ST) production in the $s$ channel, and top quark production in association with vector bosons, are simulated at NLO in QCD with \MGvATNLO using the \textsc{FxFx} merging scheme~\cite{Frederix:2012ps}. The multijet contribution, estimated with the use of samples generated at LO with the same generator, is negligible after analysis selections.

All the simulated processes use the NNPDF 3.0~\cite{Ball:2014uwa} parton distribution functions (PDFs), and are interfaced with~\PYTHIA~8.205~\cite{Sjostrand:2007gs,Sjostrand:2006za} for the parton showering and hadronization. The CUETP8M1 underlying event tune~\cite{Khachatryan:2015pea} is used in all samples, except for top quark pair production, which adopts the CUETP8M2T4 tune~\cite{CMS-PAS-TOP-16-021}.

Additional minimum bias {\Pp}{\Pp}~interactions within the same or adjacent bunch crossings (pileup) are added to the simulated processes, and events are weighted to match the observed average number of interactions per bunch crossing. Generated events are processed through a full CMS detector simulation based on {\GEANTfour}~\cite{Agostinelli:2002hh} and reconstructed with the same algorithms used for collision data.\par}

\section{Event selection}\label{sec:selection}
{\tolerance=8000
Events are classified into three independent categories ($0\ell$, $2\Pe$, and $2\mu$), based on the number and flavor of the reconstructed leptons. Events are required to have at least two jets with $\pt>30\GeV$ and $\abs{\eta}<2.4$ to be suitable candidates for the reconstruction of the $\Ph\to\bbbar$ decay. If more than two jets fulfill the requirements, the ones with the largest {\cPqb} tagging discriminator value are used to reconstruct the Higgs boson candidate. The efficiency of the correct assignment of the reconstructed jets to initial quarks originating from the Higgs boson decay varies between 80 and 97\%, after applying the event selections, depending on the category and final state.

In the $0\ell$ category, no isolated electron or muon with $\pt>10\GeV$ is allowed. Events containing isolated hadronic decays of the $\tau$ leptons with $\pt>18\GeV$ are vetoed as well. A selection is applied on the reconstructed \ptmiss, which is required to be larger than 200\GeV, such that the \ptmiss trigger is at least 95\% efficient. In order to select a topology where the \PZ boson recoils against the Higgs boson, a Lorentz boost requirement of $200\GeV$ on the \pt of the Higgs boson candidate, $\pt^{\bbbar}$, is applied.

Multijet production is suppressed by requiring that the minimum azimuthal angular separation between all jets and the missing transverse momentum vector must satisfy $\Delta\phi\text{(jet, \ptvecmiss)} > 0.4$. The multijet simulation is validated in a region obtained by inverting the $\Delta\phi$ selection, finding a good description of data.
When the \PZ boson decays to neutrinos, the resonance mass \mA cannot be reconstructed directly. In this case, \mA is estimated by computing the transverse mass from the \ptvecmiss and the four-momenta of the two jets used to reconstruct the Higgs boson candidate, defined as $\mtZH = \sqrt{\smash[b]{2 \ptmiss \pt^{\Ph}\, [1-\cos{\Delta \phi\text{(\Ph, \ptvecmiss)} }]}}$, which has to be larger than 500\GeV. The efficiency of these selections for signal events with $\mA \lesssim 500\GeV$ is small, because the \pt of the \PZ boson is not sufficient to produce a \ptmiss large enough to pass the selection; thus, the contribution of the $0\ell$ category is significant only for large \mA.

In the $2\Pe$ and $2\mu$ categories, events are required to have at least two isolated electrons or muons within the detector geometrical acceptance. The \pt threshold on the lepton is referred to as $\pt^\ell$, and is set to 30\GeV for the lepton with highest \pt, and to 10\GeV for the lepton with next-highest \pt. The \PZ boson candidate is formed from the two highest \pt, opposite charge, same-flavor leptons, and must have an invariant mass \mll between 70 and 110\GeV. The \mll selection lowers the contamination from \ttbar dileptonic decays, and significantly reduces the contribution from $\PZ\to\tau\tau$ decays. The reconstructed \ptmiss also has to be smaller than 100\GeV to reject the \ttbar background. In order to maximize the signal acceptance, no Lorentz boost requirement is applied to the \PZ and \Ph boson candidates in the dileptonic categories. The \A boson candidate is reconstructed from the invariant mass \mZH of the \PZ and \Ph boson candidates.

If the two jets originate from a Higgs boson, their invariant mass is expected to peak close to 125\GeV. Events with a dijet invariant mass \mjj between 100 and 140\GeV enter the signal regions (SRs); otherwise, if $\mjj<400\GeV$, they fall in dijet mass sidebands, which are used as control regions (CRs) to estimate the contributions of the main backgrounds. Signal regions are further divided by the number of jets passing the {\cPqb} tagging requirement (1, 2, or at least 3 {\cPqb} tags). The 3 {\cPqb} tag category has been defined to select the additional {\cPqb} quarks from {\cPqb} quark associated production. In this region, at least one additional jet, other than the two used to reconstruct the \Ph boson, has to pass the kinematic selections and {\cPqb} tagging requirements. The fraction of signal events passing the \mjj selection in the SR is 66--82\% and 45--65\% in the 1 and 2 {\cPqb} tag categories, respectively. Control regions for the \Zjets background share the same selections as the corresponding SR, except for the \mjj mass window.

Dedicated CRs are defined to estimate the \ttbar and \Wjets backgrounds, which may enter the $0\ell$ SR if the lepton originating from the \PW~decay is outside the detector geometrical acceptance or is not reconstructed. Two \Wjets CRs share the same selection as in the $0\ell$ categories, but require exactly one electron or one muon passing the same trigger and selections of the leading lepton in the $2\ell$ categories. In order to mimic the kinematics of leptonic \PW~decays, where the lepton is outside the geometrical acceptance or is not reconstructed in the detector, the \ptmiss is recalculated by removing the contribution of the lepton. The $\min(\Delta\phi)$ requirement is removed, and the dijet invariant mass selection is not applied, as the signal is absent in $1\ell$ final states. Events are required to have three or fewer jets, none of them {\cPqb} tagged, to reduce the \ttbar contribution.

Four different CRs associated with the production of events containing top quarks are defined by inverting specific selections with respect to the SR definition. Dileptonic \ttbar control regions require the same selections as the $2\Pe$ and $2\mu$ categories with two {\cPqb} tags, but the dilepton invariant mass region around the nominal \PZ boson mass is vetoed ($50<\mll<70\GeV$ or $\mll>110\GeV$), and the \mjj selection is dropped. Two additional top quark CRs are defined specifically for \ttbar events where only one of the two \PW~bosons decays into an electron or a muon, and the lepton is not reconstructed. These events contribute to the \ttbar contamination in the $0\ell$ categories. The two single-lepton top quark CRs have the same selections as the two \Wjets CRs, but in this case the jet and {\cPqb} tag vetoes are inverted to enrich the \ttbar composition.

An important feature of the signal is that the two {\cPqb} jets originate from the decay of the \Ph boson, whose mass is known with better precision than that provided by the \bbbar invariant mass resolution. The measured jet \pt values are therefore scaled according to their corresponding uncertainty given by the jet energy scale corrections to constrain the dijet
invariant mass to $\mjj=125\GeV$. The kinematic constraint on the \Ph boson mass improves the relative four-body invariant mass resolution from 5--6\% to 2.5--4.5\% for the smallest and largest values of \mA, respectively.
Similarly, in the $2\ell$ channels, the electron and muon \pt are scaled to a dilepton invariant mass $\mll = m_{\PZ}$. The effect on the \mA resolution of the kinematic constraint on the leptons is much smaller than the one of the jets, because of their better momentum resolution. 

In the $2\Pe$ and $2\mu$ categories, the \A boson decay chain yields an additional characteristic, which helps distinguish it from SM background. Five helicity-dependent angular observables fully describe the kinematics of the $\A\to\PZ\Ph\to\ell\ell\bbbar$ decay: the angle between the directions of the \PZ boson and the beam in the rest frame of the \A boson ($\cos \theta^*$); the decay angle between the direction of the negatively charged lepton relative to the \PZ boson momentum vector in the rest frame of the \PZ boson ($\cos\theta_1$), which is sensitive to the transverse polarization of the \PZ boson along its momentum vector; the angle between a jet from the \Ph boson and the \Ph boson momentum vector in the \Ph boson rest frame ($\cos\theta_2$); the angle between the \PZ and \Ph boson decay planes in the rest frame of the \A boson ($\Phi$); the angle between the \Ph boson decay plane and the plane where the \Ph boson and the beam directions lie in the \A boson rest frame ($\Phi_1$).
The discriminating power and low cross-correlation make these angles suitable as input to a likelihood ratio multivariate discriminator. This angular discriminant is defined as:
\begin{equation}
\mathcal{D} (x_1, \dots, x_N) = \frac{\displaystyle \prod_{i=1}^{N} s_i (x_i)}{\displaystyle \prod_{i=1}^{N} s_i (x_i)+\prod_{i=1}^{N} b_i (x_i)}
\end{equation}
where the index $i$ runs from $1$ to $5$ and corresponds to the number $N$ of angular variables $x_i$, and $s_i$ and $b_i$ are the signal and \Zjets background probability density functions of the $i$-th variable, respectively. A selection of $\mathcal{D}>0.5$ is applied in all $2\Pe$ and $2\mu$ SRs and CRs, except those with three {\cPqb} tags due to the low event count. This working point retains 80\% of the signal efficiency and rejects 50\% of the \Zjets background.

Considering that top quark pair production may be as large as 50\% of the total background in certain regions of the parameter space, a second likelihood ratio discriminator is built specifically to reject the \ttbar events. This discriminator uses only the \mll and \ptmiss variables. The background probability density function considers only the top quark background in order to achieve the maximum separation between events with a genuine leptonically decaying \PZ boson recoiling against a pair of jets and the more complex topologies such as \ttbar decays. Selecting events with a discriminator output larger than 0.5 rejects 75\% of the \ttbar events with a signal efficiency of 85\%. This selection is applied to the dileptonic SRs and to the \Zjets CRs.

The SRs and CRs selections are summarized in Table~\ref{tab:selections}. The product of the signal acceptance and selection efficiency as a function of \mA is presented in Fig.~\ref{fig:efficiency} separately for the gluon-gluon fusion and {\cPqb} quark associated production modes.

\begin{table*}[!htb]
  \centering
  \topcaption{Definition of the signal and control regions. In $2\ell$ regions, the leptons are required to have opposite electric charge. The entries marked with $\dagger$ indicate that the \ptmiss is calculated subtracting the four momentum of the lepton.}
  \label{tab:selections}
  \cmsTable{
  \begin{tabular}{lccccccc}
    \hline
    Region            & $0\ell$ SR & $0\ell$ $\PZ$ CR & $1\ell$ $\PW$ CR & $1\ell$ {\cPqt} CR & $2\ell$ SR & $2\ell$ $\PZ$ CR & $2\ell$ {\cPqt} CR \\
    \hline
    Leptons           & \multicolumn{2}{c}{$\Pe,\mu,\tau$ veto} & \multicolumn{2}{c}{$1\Pe$ or $1\mu$} & \multicolumn{3}{c}{$2\Pe$ or $2\mu$} \\
    $\pt^\ell$ (\GeV) & \multicolumn{2}{c}{\NA} & \multicolumn{2}{c}{${>}55$} & \multicolumn{3}{c}{${>}55,20$} \\
    \mll (\GeV)       & \NA & \NA & \NA & \NA & \multicolumn{2}{c}{$70{<} \mll {<}110$} & ${<}70, {>}110$ \\
    \ptmiss (\GeV)    & ${>}200$ & ${>}200$ & ${>}200^\dagger$ & ${>}200^\dagger$ & ${<}100$ & ${<}100$ & \NA \\
    Jets              & ${\geq}2$ or 3 & ${\geq}2$ & ${\leq}3$ & ${\geq}4$ & ${\geq}2$ or 3 & ${\geq}2$ & ${\geq}2$ \\
    {\cPqb}-tagged jets   & 1, 2, or 3 & 0, 1, 2, or 3 & 0 & ${\geq}1$ & 1, 2, or 3 & 0, 1, 2, or 3 & ${\geq}2$ \\
    $\pt^{\bbbar}$ (\GeV) & ${>}200$ & ${>}200$ & ${>}200$ & ${>}200$ & \NA & \NA & \NA \\
    \mjj (\GeV)       & ${>}100,{<}140$ & ${<}100,{>}140$ & \NA & \NA & ${>}100,{<}140$ & ${<}100,{>}140$ & \NA \\
    $\Delta \varphi(\text{j},\ptvecmiss)$ & ${<}0.4$ & ${<}0.4$ & \NA & \NA & \NA & \NA & \NA \\
    Angular $\mathcal{D}$ & \NA & \NA & \NA & \NA & ${>}0.5$ & ${>}0.5$ & \NA \\
    Top quark $\mathcal{D}$ & \NA & \NA & \NA & \NA & ${>}0.5$ & ${>}0.5$ & \NA \\
    \hline
  \end{tabular}
  }
\end{table*}

\begin{figure*}[!htb]\centering
    \includegraphics[width=0.495\textwidth]{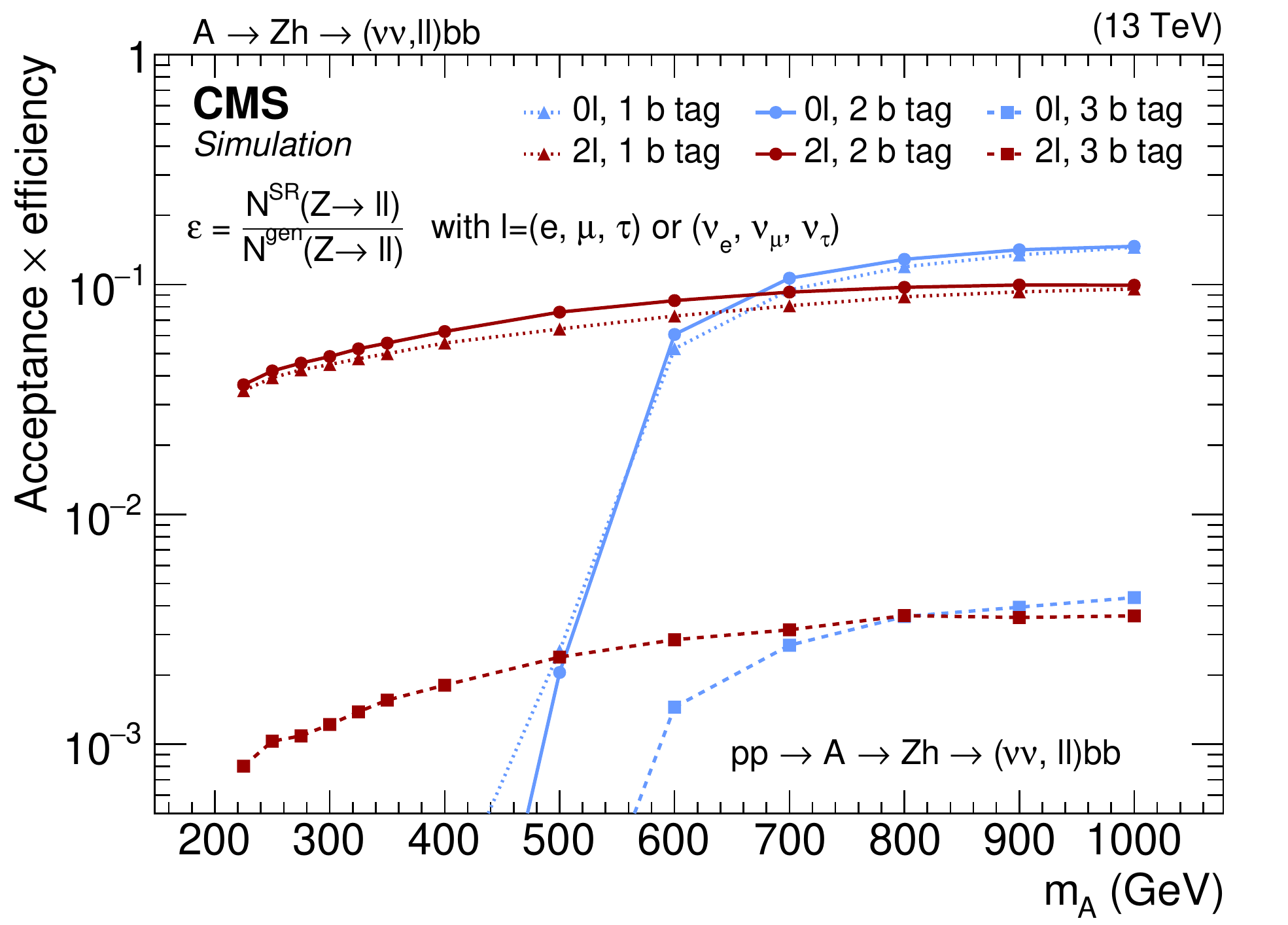}
    \includegraphics[width=0.495\textwidth]{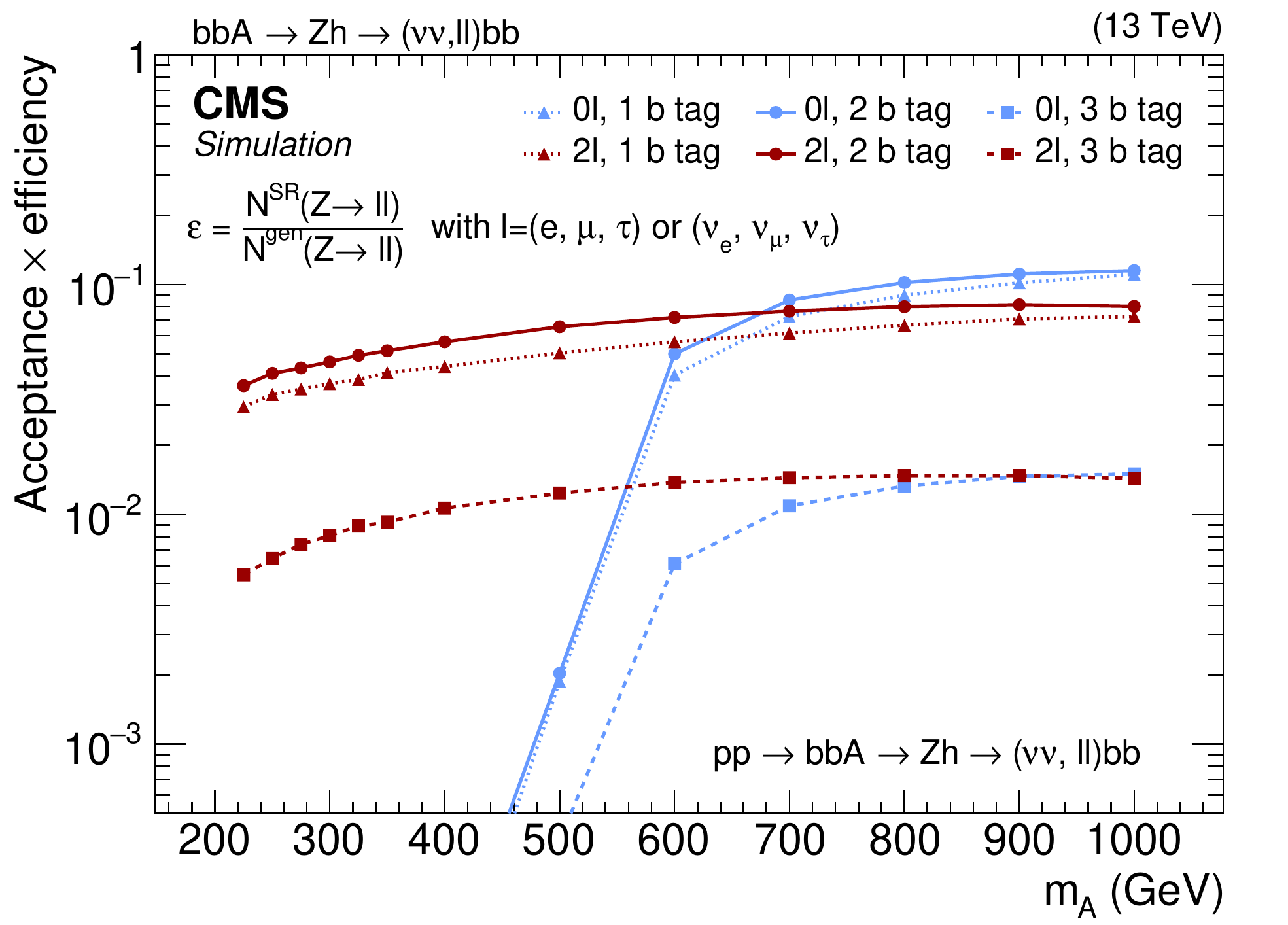}
  \caption{Product of the signal acceptance and selection efficiency $\varepsilon$ for an \A boson produced via gluon-gluon fusion (left) and in association with {\cPqb} quarks (right) as a function of \mA. The number of events passing the signal region selections is denoted as $N^\mathrm{SR}$, and $N^\mathrm{gen}$ is the number of events generated before applying any selection.}
  \label{fig:efficiency}
\end{figure*}
\par}

\section{Systematic uncertainties}\label{sec:syst}

The uncertainties in the trigger efficiency and the electron, muon, and $\tau$ lepton reconstruction, identification, and isolation efficiencies are evaluated through studies of events with dilepton invariant mass around the \PZ boson mass, and the variation of the event yields with respect to the expectation from simulation amount to approximately 2--3\% for the categories with charged leptons, and 1\% in the $0\ell$ categories~\cite{Khachatryan:2015hwa,Chatrchyan:2012xi,Khachatryan:2015dfa}. The impact of the lepton energy and momentum scale and resolution is small after the kinematic constraint on \mll. 
The jet energy scale and resolution~\cite{Khachatryan:2016kdb} affect both the selection efficiencies and the shape of the \ptmiss and \mtZH distributions, and are negligible in the $2\ell$ channels after the kinematic constraint on the dijet mass has been applied. The jet four-momentum is varied by the corresponding uncertainties, and the effect is propagated to the final distributions. The jet energy scale is responsible for a 2--6\% variation in the numbers of background and signal events; the jet energy resolution contributes an additional 1--2\% uncertainty. The effects of jet energy scale and resolution uncertainties, as well as the energy variation of the unclustered objects in the event, are propagated to the \ptmiss and \mtZH distributions.
The {\cPqb} tagging uncertainty~\cite{BTV-16-002} in the signal yield depends on the jet \pt and thus on the mass of the resonance, and the impact on the event yield ranges from 2 to 4\% in the 1 {\cPqb} tag category, 4 to 8\% in the 2 {\cPqb} tag category, and 8 to 12\% in the 3 {\cPqb} tag category.

The signal and background event yields are affected by the uncertainties on the choice of PDFs~\cite{Butterworth:2015oua} and the factorization and renormalization scale uncertainties.
The former are derived with \textsc{SysCalc}~\cite{Kalogeropoulos:2018cke}, and the latter are estimated by varying the corresponding scales up and down by a factor of two~\cite{Catani:2003zt}. The effect of both these uncertainties can be as large as 30\% depending on the generated signal mass. The effect of the PDF uncertainties on the signal and background lepton acceptance is estimated to be an average of 3\% per lepton. The top quark background is also affected by the uncertainty associated with the simulated \pt spectrum of top quarks~\cite{Khachatryan:2016mnb}, which results in up to a 14\% yield uncertainty. The \Vjets backgrounds are affected by the uncertainties on the QCD and EW NLO corrections, as described in Section~\ref{sec:mcsimulation}.

A systematic uncertainty is assigned to the interpolation between the two mass sidebands to the SR, defined as the difference in the ratio between data and simulated background in the lower and upper sidebands, and ranges between 2 and 10\% depending on the channel. The extrapolation to the 3 {\cPqb} tag regions is covered by a large uncertainty (20--46\%) assigned to the overall background normalization, and derived by taking the ratio between data and the simulation in the 3 {\cPqb} tag control regions.
In the dilepton categories, a dedicated uncertainty is introduced to cover for minor mismodeling effects. The background distribution is reweighted with a linear function of the event centrality (defined as the ratio between the sums of the \pt and the energy of the two leptons and two jets in the rest frame of the four objects) in all simulated events, and the effect is propagated to the \mZH distributions as a systematic uncertainty.

Additional systematic uncertainties affect the event yields of backgrounds and signal come from pileup contributions and integrated luminosity~\cite{CMS:lumi}. The uncertainty from the limited number of simulated events is treated as in Ref.~\cite{Barlow1993219}.
A summary of the systematic uncertainties is reported in Table~\ref{tab:Sys}.

\begin{table*}[!htb]
  \centering
  \topcaption{Summary of statistical and systematic uncertainties for backgrounds and signal. The uncertainties marked with $\checkmark$ are also propagated to the \mZH and \mtZH distributions.}
  \label{tab:Sys}
  \cmsTable{
  \begin{tabular}{lccccc}
    \hline
    & \multirow{2}{*}{Shape} & Main backgrounds & Other backgrounds & Signal \\
    & & (\Vjets, \ttbar) & ($\ST$, $\VV$, $\V\Ph$) & \\
    \hline
    Lepton and trigger          & $\checkmark$ & \multirow{2}{*}{\NA} &\multirow{2}{*}{2--3\%} & \multirow{2}{*}{2--3\%} \\
    \quad efficiency \\
    Jet energy scale            & $\checkmark$ &           \NA & 5\% & 2--6\% \\
    Jet energy resolution       & $\checkmark$ &           \NA & 2\% & 1--2\% \\
    {\cPqb} tagging             & $\checkmark$ &           \NA & 4\% & 4--12\% \\
    Unclustered \ptmiss         & $\checkmark$ &           \NA & 1\% & 1\% \\
    Pileup                      & $\checkmark$ &           \NA & 1\% & 1\% \\
    PDF                         & $\checkmark$ &           \NA & 3--5\% & 4--8\%\\
    Top quark \pt modeling      & $\checkmark$ & 8--14\% (only \ttbar) & \NA & \NA \\
    Fact. and renorm. scale     & $\checkmark$ &           \NA & 2--6\% & 6--14\% \\
    Monte Carlo modeling        & $\checkmark$ & \multicolumn{2}{c}{1--15 \%} & \NA \\
    Monte Carlo event count     & $\checkmark$ & \multicolumn{2}{c}{1--20\%} & \NA \\
    Interpolation to SR         &              & \multicolumn{2}{c}{2--10\%} & \NA \\
    Extrapolation to ${\geq}3$ {\cPqb} tag SR &       & \multicolumn{2}{c}{20--46\% (${\geq}3$ {\cPqb} tag only)} & \NA \\
    Cross section               &              &           \NA & 2--10\% & \NA     \\
    Integrated luminosity       &              &           \NA & 2.5\% & 2.5\%    \\
    \hline
  \end{tabular}
  }
\end{table*}

\section{Results and interpretation}\label{sec:results}

\begin{figure}[!htb]\centering
  \includegraphics[width=.5\textwidth]{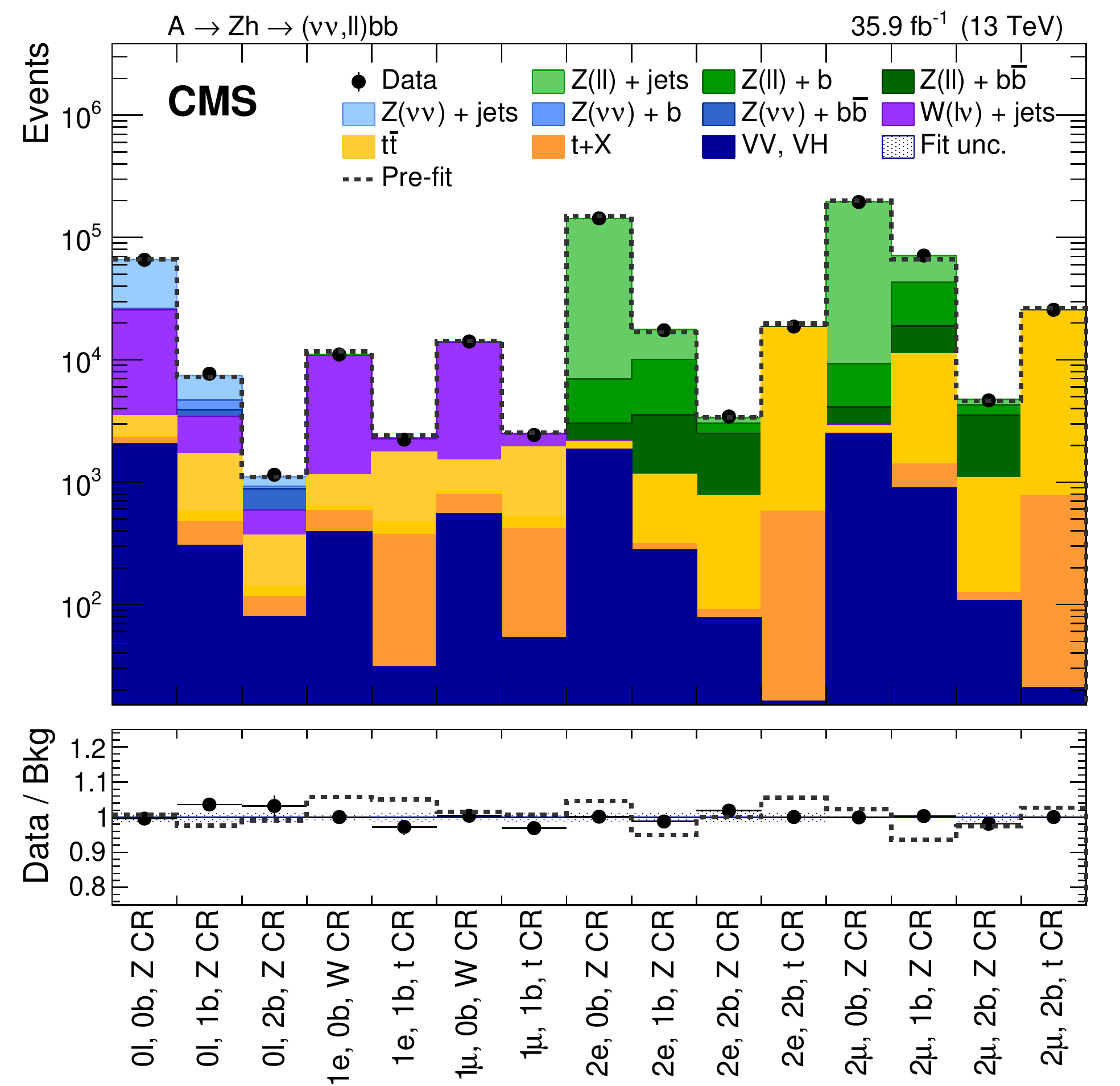}
  \caption{Pre- (dashed gray lines) and post-fit (stacked histograms) numbers of events in the different control regions used in the fit. The label in each bin summarizes the control region definition, the selection on the number and flavor of the leptons, and the number of {\cPqb}-tagged jets. The bottom panel depicts the ratio between the data and the SM backgrounds.}
  \label{fig:fit_CR}
\end{figure}

\begin{figure*}[!hbtp]\centering
  \includegraphics[width=0.43\textwidth]{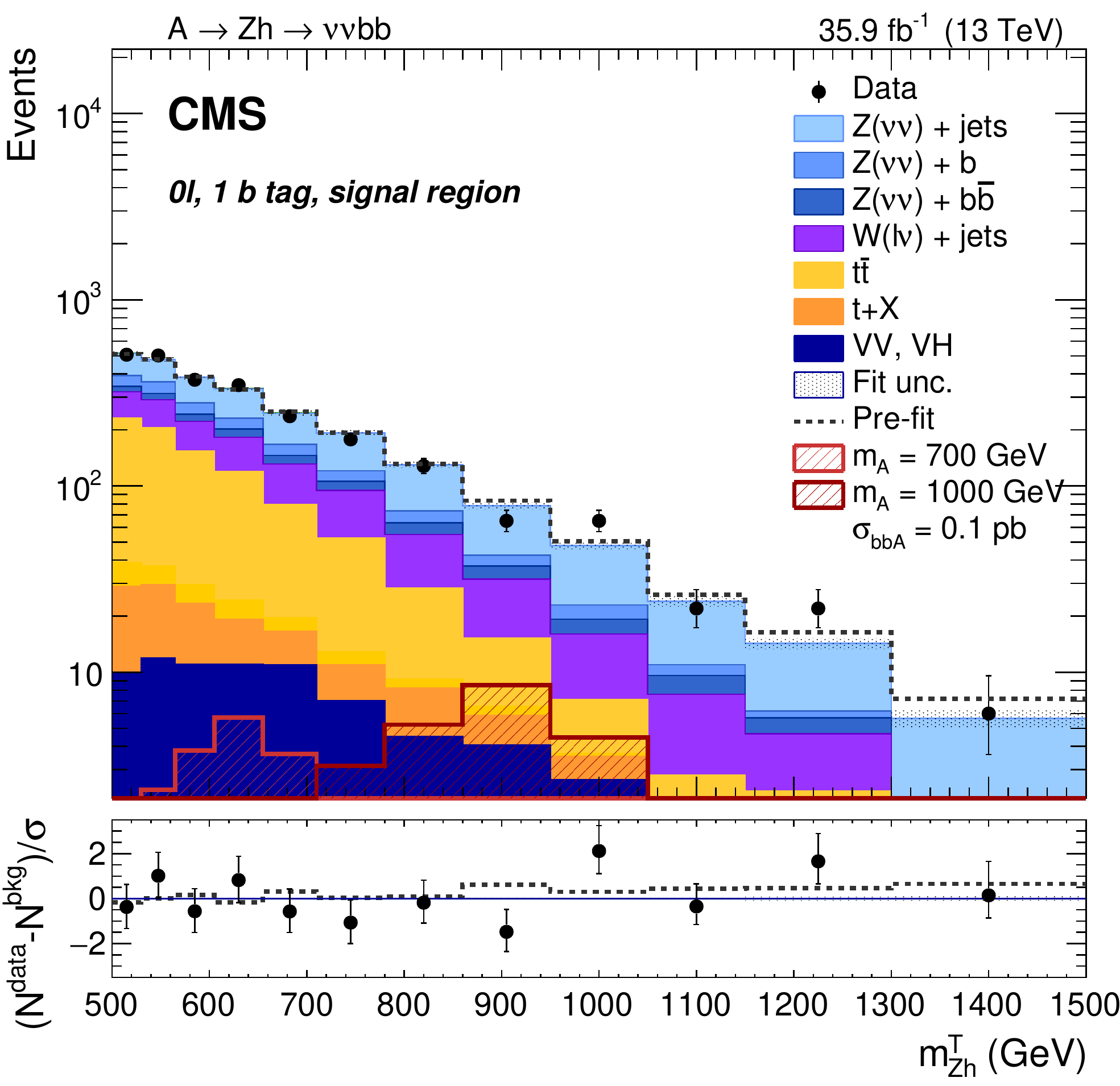}
  \includegraphics[width=0.43\textwidth]{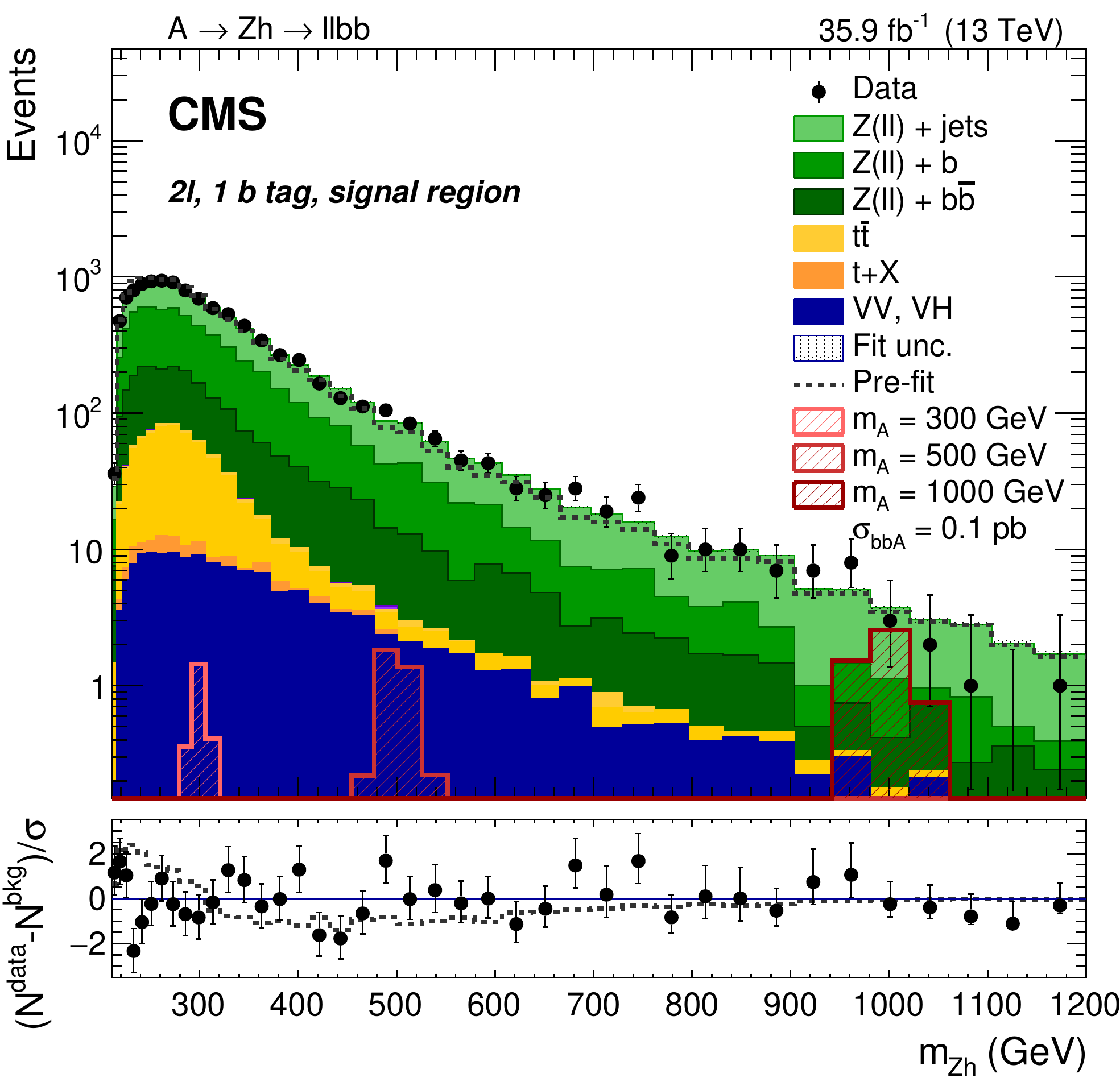}

  \includegraphics[width=0.43\textwidth]{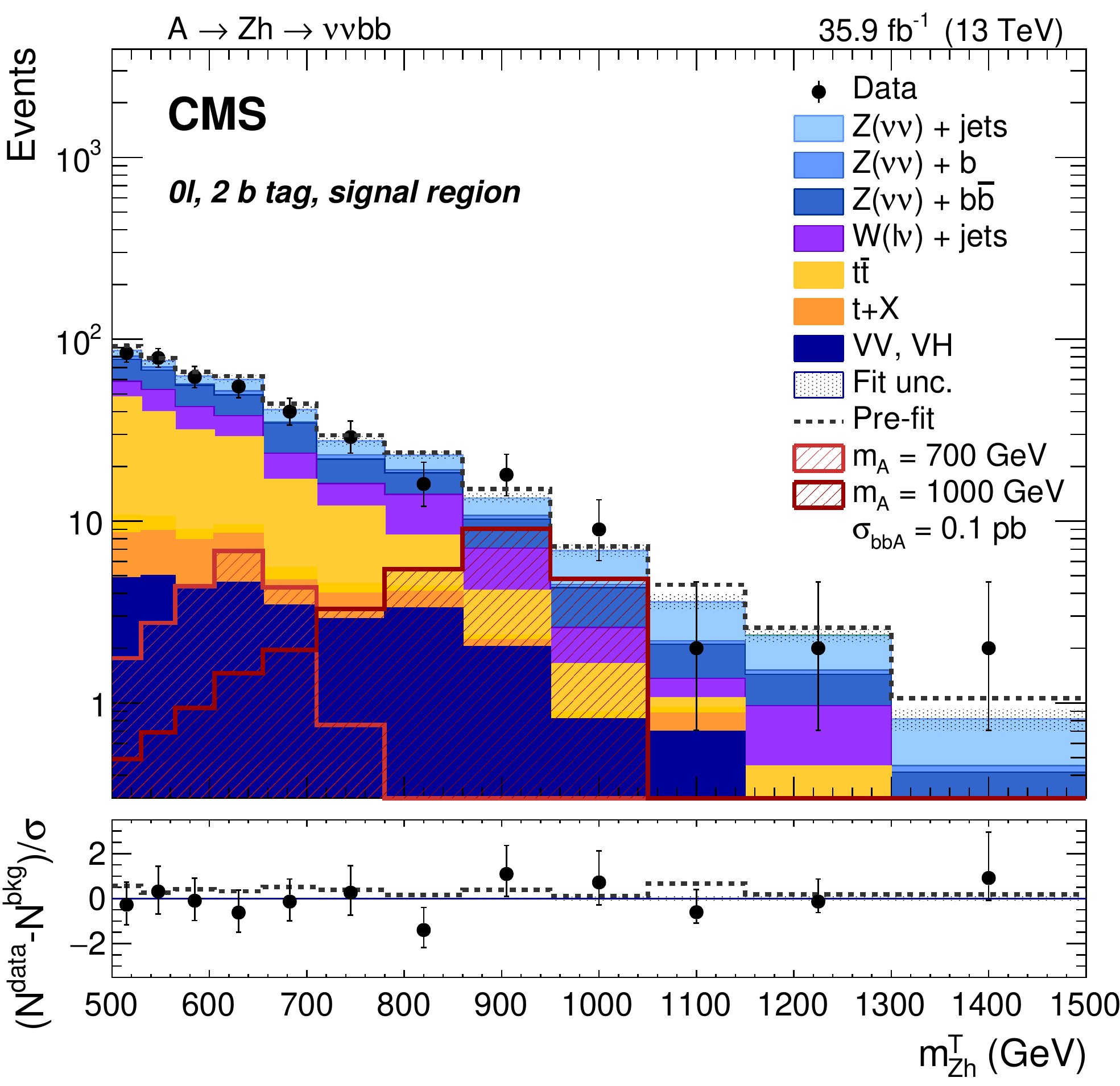}
  \includegraphics[width=0.43\textwidth]{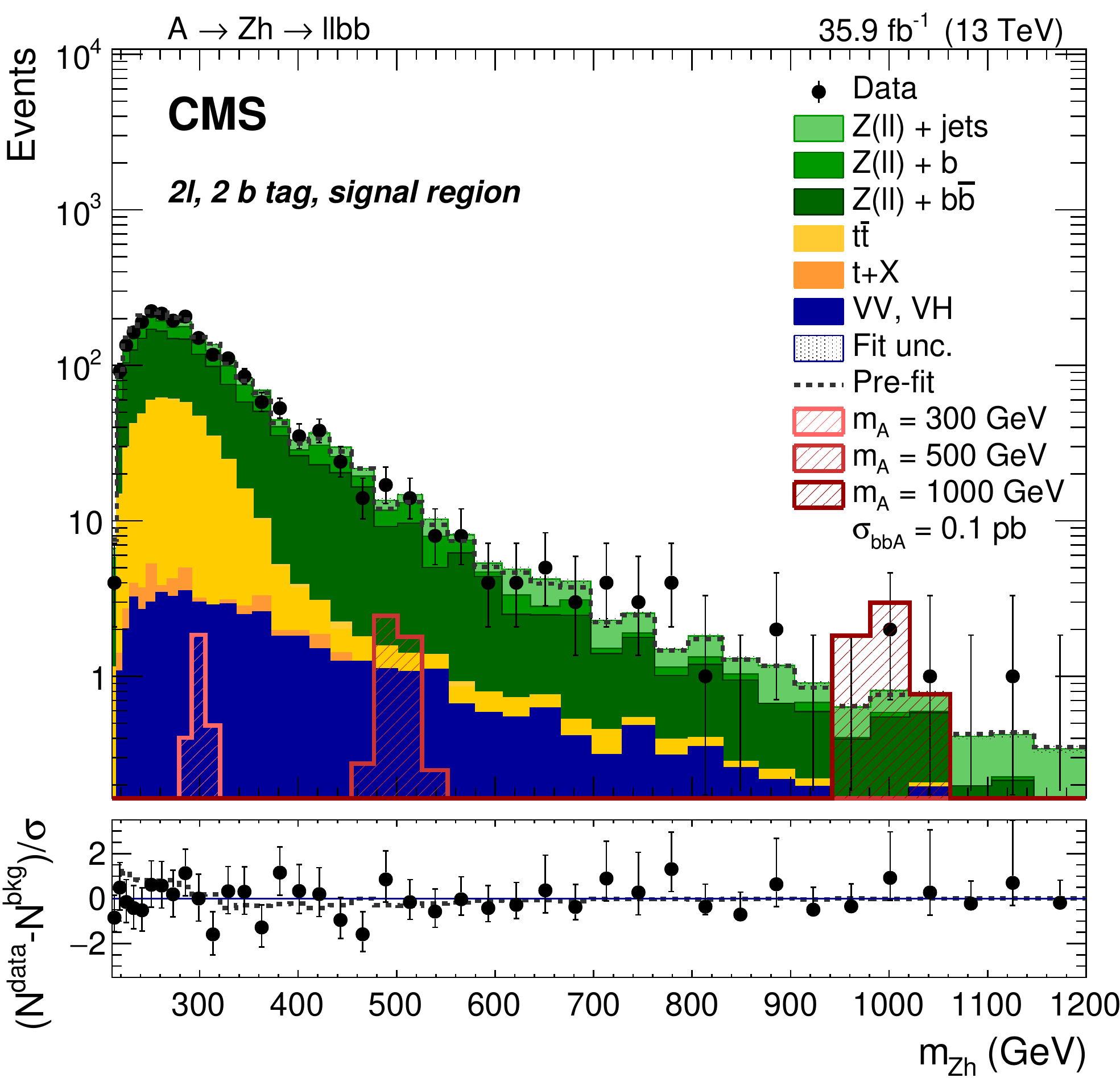}

  \includegraphics[width=0.43\textwidth]{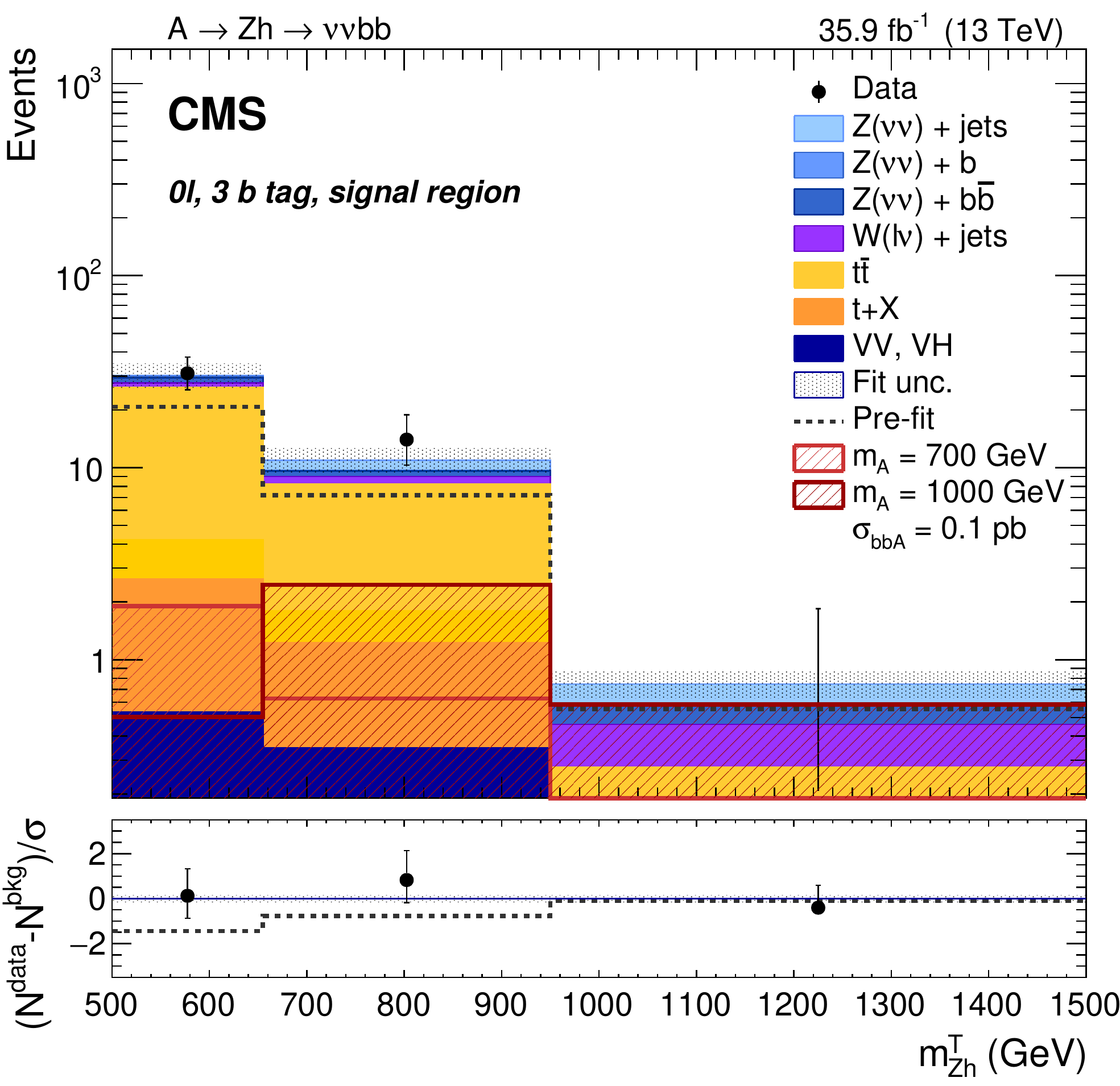}
  \includegraphics[width=0.43\textwidth]{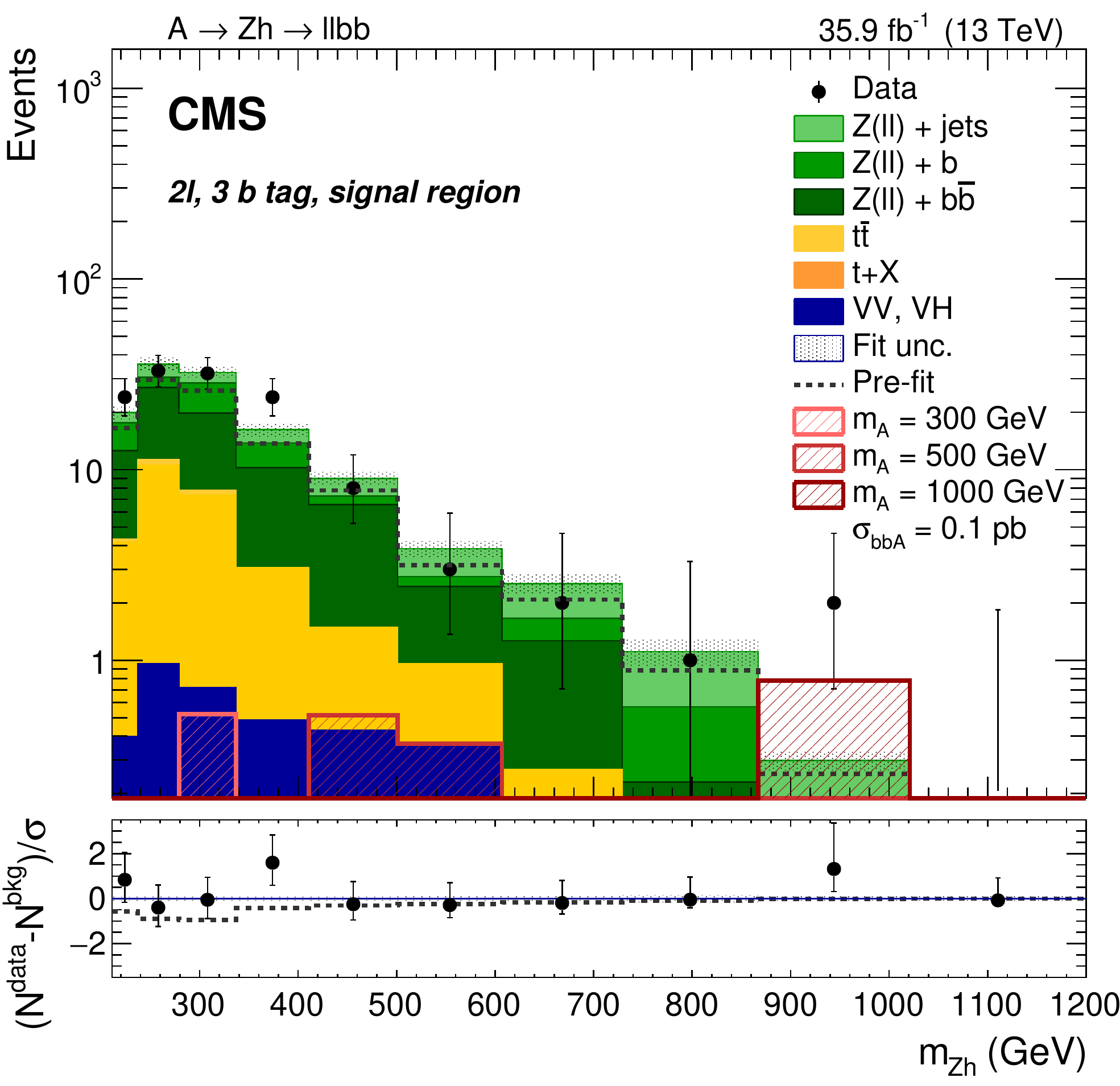}

  \caption{Distributions of the \mtZH variable in the $0\ell$ categories (left) and \mZH in the $2\ell$ categories (right), in the 1 {\cPqb} tag (upper), 2 {\cPqb} tag (center), and 3 {\cPqb} tag (lower) SRs. In the $2\ell$ categories, the contribution of the $2\Pe$ and $2\mu$ channels have been summed. The gray dotted line represents the sum of the background before the fit; the shaded area represents the post-fit uncertainty. The hatched red histograms represent signals produced in association with {\cPqb} quarks and corresponding to $\sigma_{\A}\B(\A\to\PZ\Ph)\B(\Ph\to\bbbar)=0.1~\unit{pb}$. The bottom panels depict the pulls in each bin, $(N^\text{data}-N^\text{bkg})/\sigma$, where $\sigma$ is the statistical uncertainty in data.
  }
  \label{fig:fit}
\end{figure*}

The signal search is carried out by performing a combined signal and background maximum likelihood fit to the number of events in the CRs, and the binned \mZH or \mtZH distributions in the SRs. Systematic uncertainties are treated as nuisance parameters and are profiled in the statistical interpretation~\cite{CLS1,CLS2,CMS-NOTE-2011-005}. The asymptotic approximation~\cite{Asymptotic} of the modified frequentist \CLs criterion~\cite{CLS1,CLS2} is used to determine limits on the signal cross section at 95\% confidence level (\CL). The background-only hypothesis is tested against the combined signal+background hypothesis in the nine categories, split according to the number and flavor of the leptons and number of {\cPqb}-tagged jets. The normalizations of the main backgrounds (\Zjets, $\PZ$+{\cPqb}, $\PZ$+\bbbar, \ttbar, $\PW$+jets) are allowed to float in the fit, and are constrained in the CRs. The multiplicative scale factors for the main backgrounds determined by the fit are reported in Table~\ref{tab:Fit_CR}, and the overall event yields in the CRs are shown in Fig.~\ref{fig:fit_CR} before and after the fit. 
The expected and observed number of events in the SRs are reported in Table~\ref{tab:Fit_SR}, and the \mZH and \mtZH distributions are shown in Fig.~\ref{fig:fit}.

\begin{table}[!htb]
  \centering
  \topcaption{Scale factors for the main backgrounds, as derived by the combined fit in the background-only hypothesis, with respect to the event yield from simulated samples.}
  \label{tab:Fit_CR}
  \begin{tabular}{lcccccccc}
    \hline
    Background      & Scale factor \\
    \hline
    $\PZ$+jets      & 0.993 $\pm$ 0.018 \\
    $\PZ$+{\cPqb}   & 1.214 $\pm$ 0.021 \\
    $\PZ$+\bbbar    & 1.007 $\pm$ 0.025 \\
    \ttbar          & 0.996 $\pm$ 0.014 \\
    $\PW$+jets      & 0.980 $\pm$ 0.023 \\
    \hline
  \end{tabular}
\end{table}

\begin{table*}[!htb]
  \centering
  \topcaption{Expected and observed event yields after the fit in the signal regions. The dielectron and dimuon categories are summed together. The ``--'' symbol represents backgrounds with no simulated events passing the selections. The signal yields refer to pre-fit values corresponding to a cross section multiplied by $\B(\A\to\PZ\Ph) \, \B(\Ph\to\bbbar)$ of $0.1\unit{pb}$ (gluon-gluon fusion for $\mA=300\GeV$, and in association with {\cPqb} quarks for $\mA=1000\GeV$). }
  \label{tab:Fit_SR}
  \cmsTable{
  \begin{tabular}{lcccccccc}
  \hline
Signal region          & $0\ell$, 1 {\cPqb} tag & $0\ell$, 2 {\cPqb} tag & $0\ell$, $\geq$3 {\cPqb} tag & $2\ell$, 1 {\cPqb} tag & $2\ell$, 2 {\cPqb} tag & $2\ell$, $\geq$3 {\cPqb} tag \\
  \hline
Data                 	& $2452 \pm 50$ & $398 \pm 20$ & $45 \pm 7$ & $10\,512 \pm 103$ & $2188 \pm 47$ & $129 \pm 11$ \\[\cmsTabSkip]
$\PZ$+jets            	& $740 \pm 12$ & $48 \pm 1$ & $2.0 \pm 0.2$ & $4118 \pm 15$ & $175 \pm 1$ & $18 \pm 1$ \\
$\PZ$+{\cPqb}        	& $220 \pm 6$ & $13 \pm 1$ & $0.46 \pm 0.06$ & $4127 \pm 18$ & $365 \pm 3$ & $23 \pm 1$ \\
$\PZ$+\bbbar          	& $134 \pm 3$ & $86 \pm 2$ & $2.5 \pm 0.3$ & $1547 \pm 11$ & $1113 \pm 7$ & $51 \pm 2$ \\
\ST                  	& $74 \pm 3$ & $18 \pm 1$ & $3.0 \pm 0.4$ & $25 \pm 0$ & $10.0 \pm 0.1$ & -\\
\ttbar               	& $750 \pm 12$ & $143 \pm 3$ & $31 \pm 3$ & $592 \pm 3$ & $473 \pm 3$ & $26 \pm 1$ \\
$\VV$, $\V\Ph$       	& $76 \pm 2$ & $32 \pm 1$ & $0.93 \pm 0.11$ & $139 \pm 1$ & $53 \pm 1$ & $3.5 \pm 0.1$ \\
$\PW$+jets            	& $458 \pm 13$ & $65 \pm 3$ & $2.4 \pm 0.3$ & $3.7 \pm 0.1$ & \NA & \NA \\[\cmsTabSkip]
Total bkg.           	& $2451 \pm 26$ & $405 \pm 8$ & $42 \pm 5$ & $10\,552 \pm 35$ & $2189 \pm 12$ & $121 \pm 3$ \\
Pre-fit bkg.         	& $2467 \pm 26$ & $427 \pm 8$ & $28 \pm 5$ & $10\,740 \pm 35$ & $2250 \pm 12$ & $100 \pm 3$ \\[\cmsTabSkip]
$\mA=300\GeV$          & \NA & \NA & \NA & $3.1 \pm 0.2$ & $3.3 \pm 0.2$ & $0.10 \pm 0.01$ \\
$\mA=1000\GeV$         & $27.3 \pm 5.2$ & $28.6 \pm 5.4$ & $3.5 \pm 0.7$ & $5.4 \pm 1.0$ & $6.1 \pm 1.2$ & $1.2 \pm 0.2$ \\
    \hline
  \end{tabular}
  }
\end{table*}

\begin{figure*}[!htb]\centering
    \includegraphics[width=0.495\textwidth]{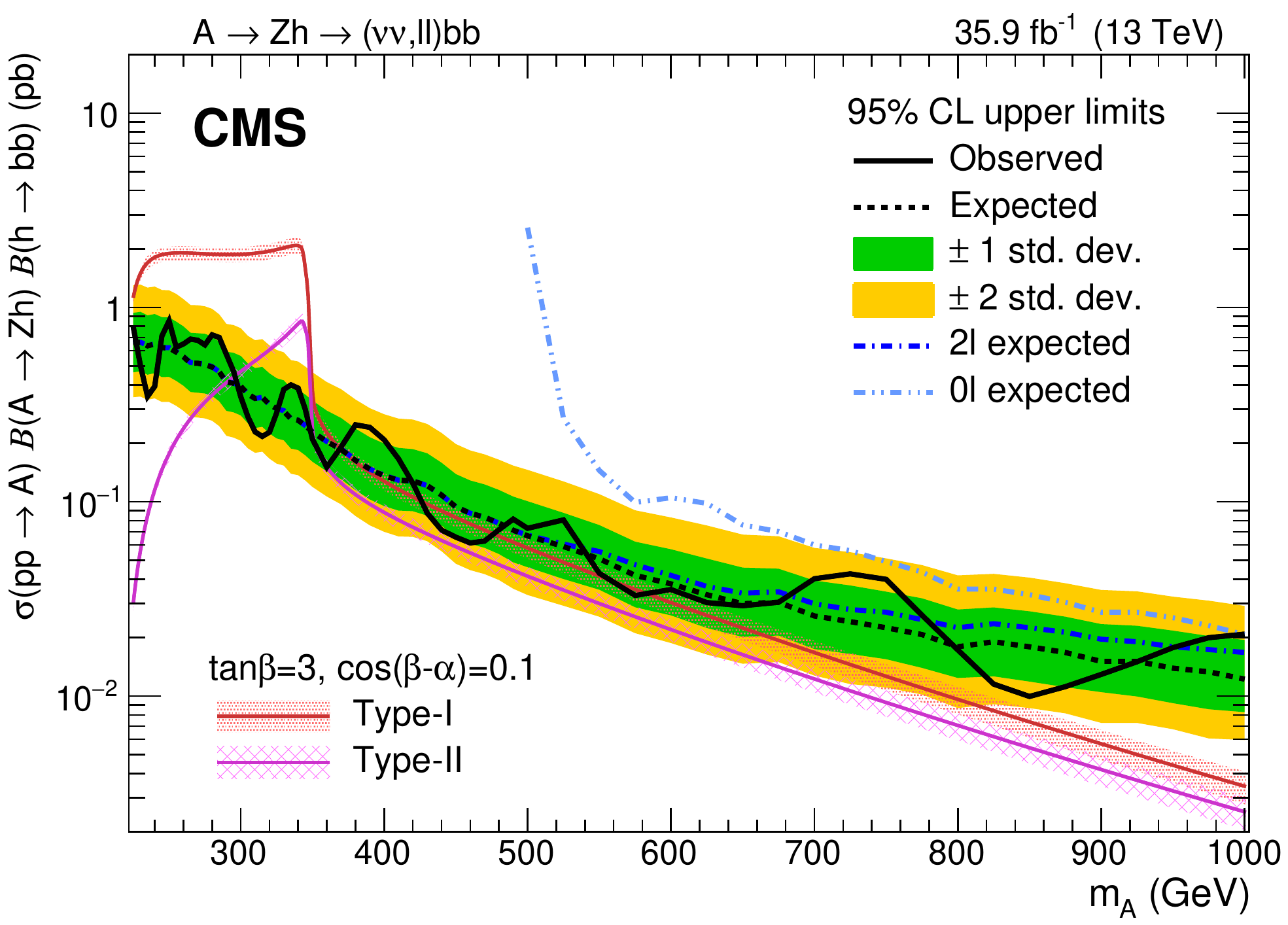}
    \includegraphics[width=0.495\textwidth]{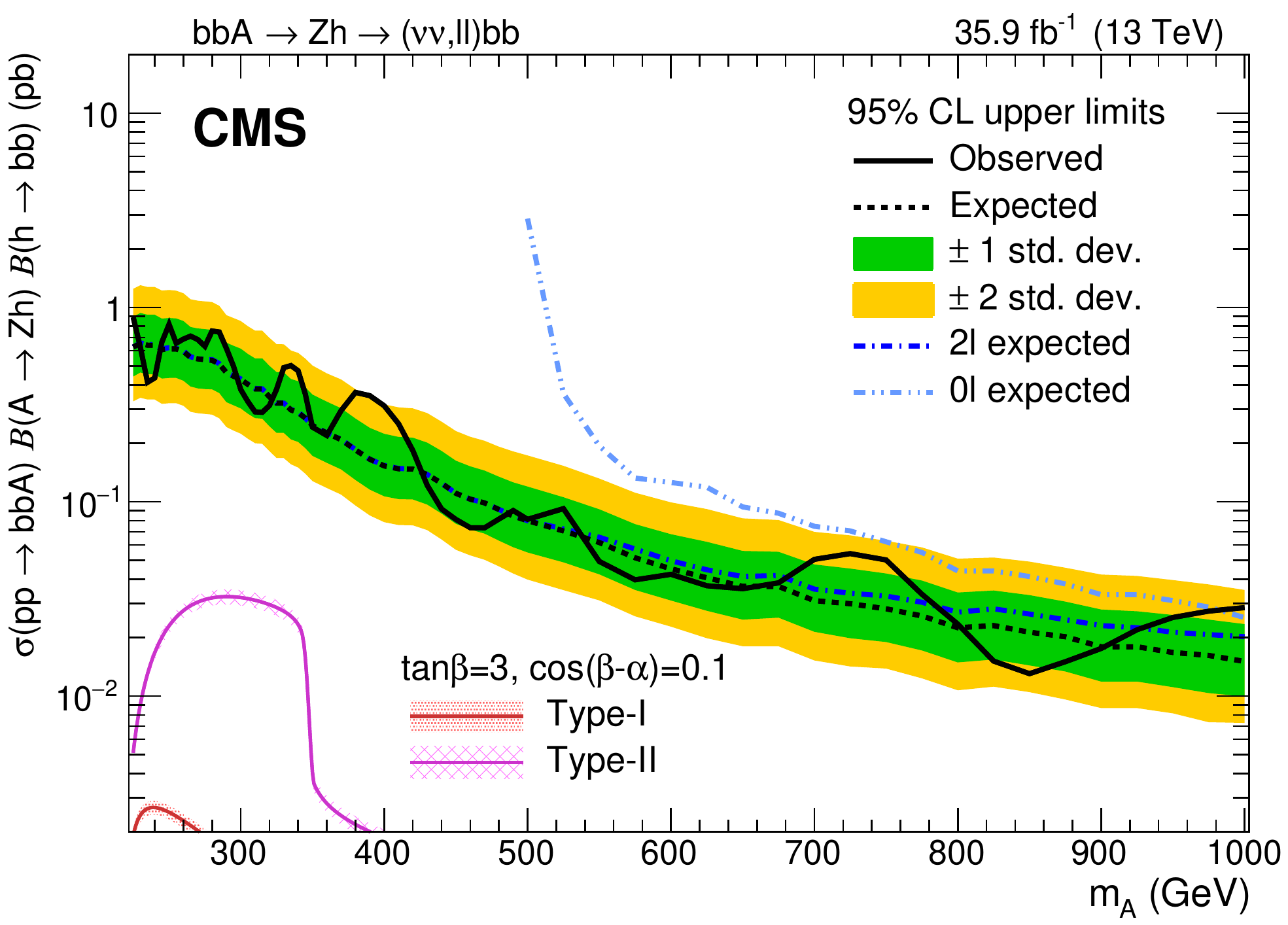}
  \caption{Observed (solid black) and expected (dotted black) 95\% \CL upper limits on $\sigma_\A \,\B(\A\to\PZ\Ph)\,\B(\Ph\to\bbbar)$ for an \A boson produced via gluon-gluon fusion (left) and in association with {\cPqb} quarks (right) as a function of \mA.
  The blue dashed lines represent the expected limits of the $0\ell$ and $2\ell$ categories separately.
  The red and magenta solid curves and their shaded areas correspond to the product of the cross sections and the branching fractions and the relative uncertainties predicted by the 2HDM Type-I and Type-II for the arbitrary parameters $\tan\beta=3$ and $\cosba=0.1$.
      }
  \label{fig:limit}
\end{figure*}

The data are well described by the SM processes. Upper limits are derived on the product of the cross section for a heavy pseudoscalar boson \A and the branching fractions for the decays $\A\to\PZ\Ph$ and $\Ph\to\bbbar$.
The limits are obtained by considering the \A boson produced via the gluon-gluon fusion and {\cPqb} quark associated production processes separately, in the approximation where the natural width of the \A boson $\Gamma_\A$ is smaller than the experimental resolution, and are reported in Fig.~\ref{fig:limit}.
An upper limit at 95\% \CL on the number of signal events is set on $\sigma_\A \,\B(\A\to\PZ\Ph) \, \B(\Ph\to\bbbar)$, excluding above 1\unit{pb} for \mA near the kinematic threshold, ${\approx}0.3\unit{pb}$ for $\mA\approx 2 m_{\PQt}$, and as low as 0.02\unit{pb} at the high end ($1000\GeV$) of the considered mass range. The sensitivity of the analysis is limited by the amount of data, and not by systematic uncertainties.
These results extend the search for a 2HDM pseudoscalar boson \A for mass up to 1\TeV, which is a kinematic region previously unexplored by CMS in the 8\TeV data analysis~\cite{Khachatryan:2015lba}. When \mA is larger than 1\TeV, the CMS analysis with merged jets~\cite{Sirunyan:2018qob} retains a better sensitivity.
The sensitivity is comparable to the ATLAS search~\cite{Aaboud:2017cxo}, which observed a mild local (global) excess of 3.6 (2.4) standard deviations corresponding to $\mA\approx 440\GeV$ in final states with $2\mu$ and 3 or more {\cPqb}-tagged jets. A slight deficit is observed by CMS in the corresponding region.

\begin{figure*}[!htb]\centering
    \includegraphics[width=0.495\textwidth]{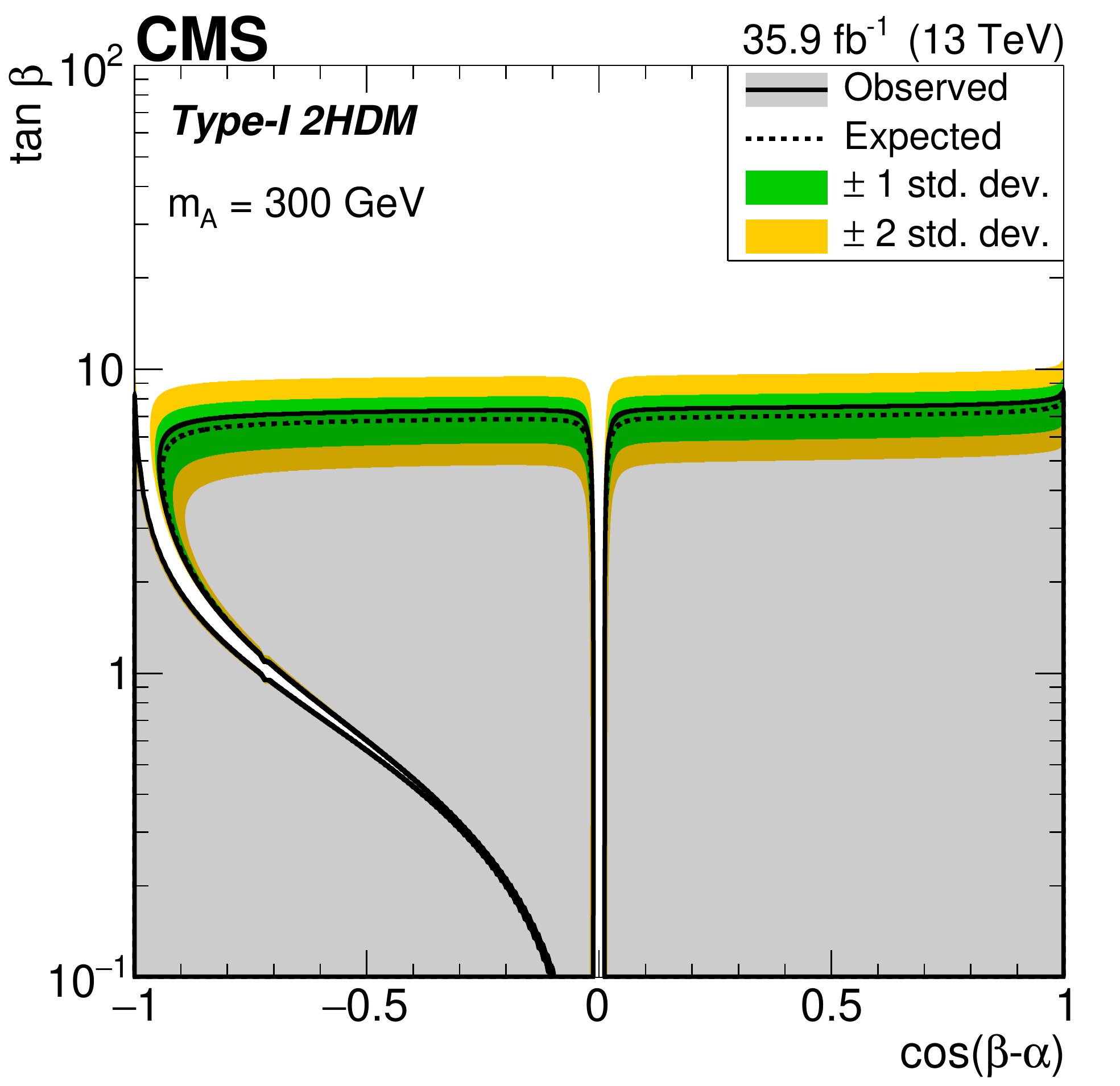}
    \includegraphics[width=0.495\textwidth]{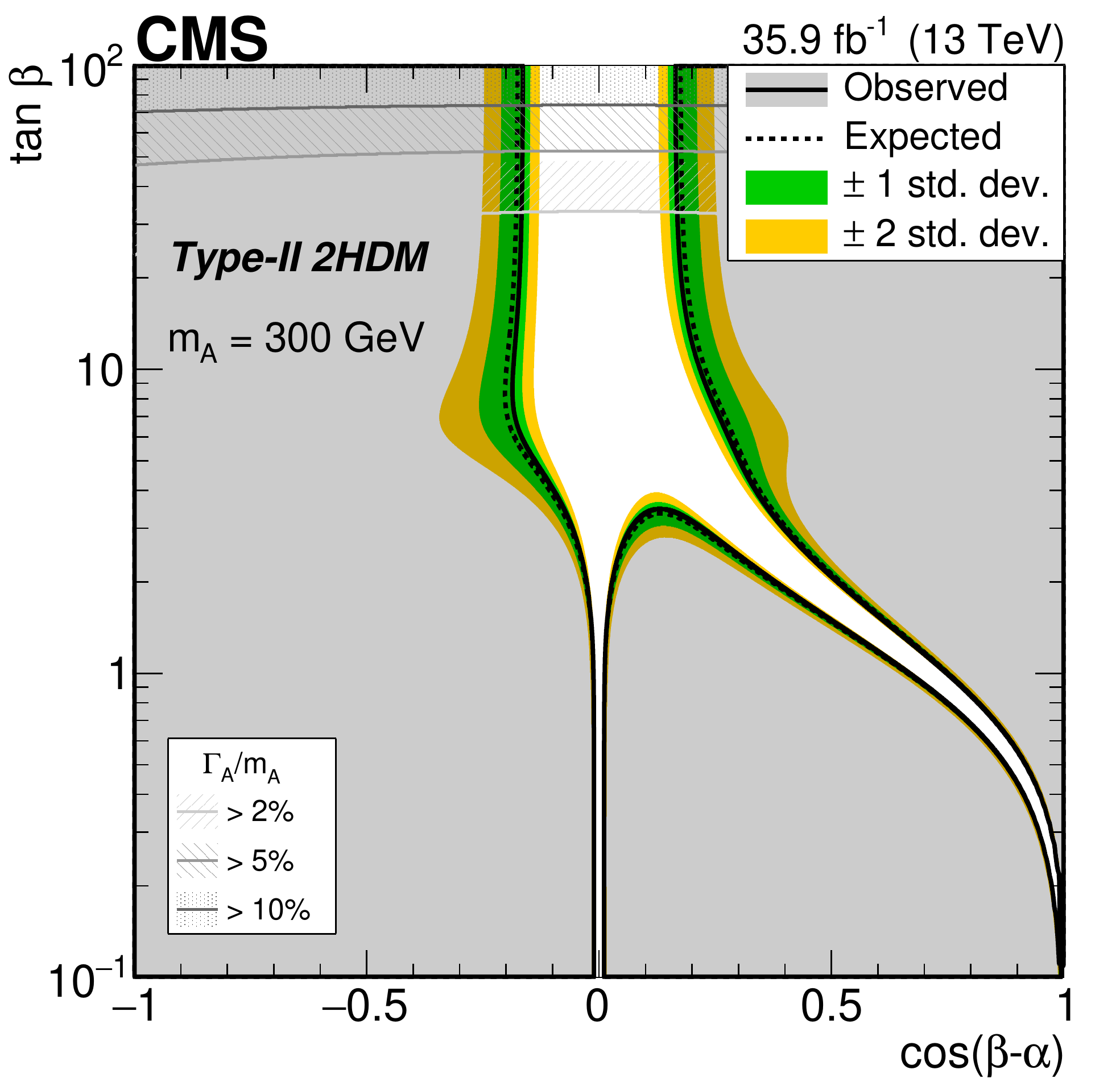}

    \includegraphics[width=0.495\textwidth]{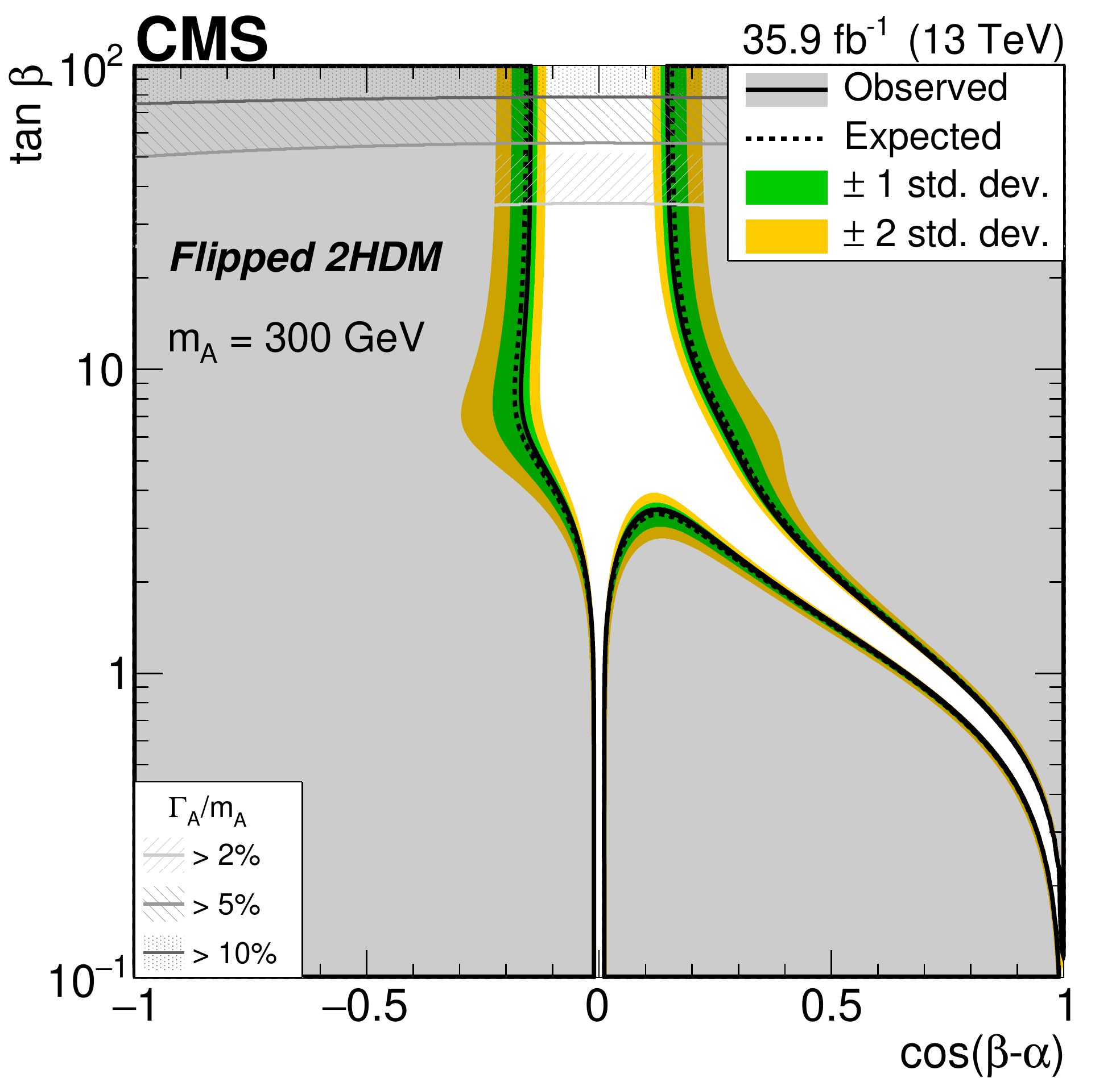}
    \includegraphics[width=0.495\textwidth]{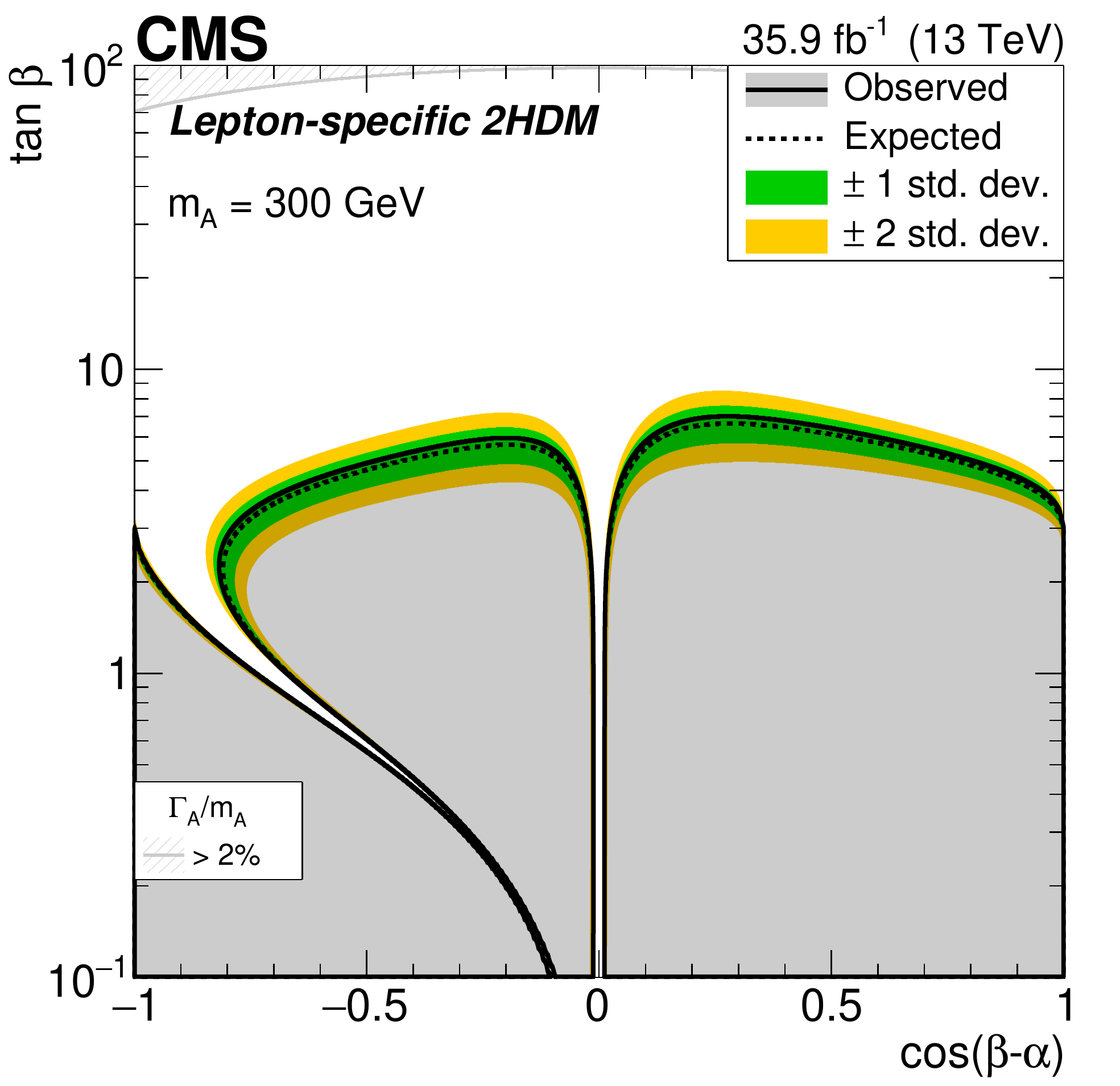}

   \caption{Observed and expected (with ${\pm}1,\,{\pm}2$ standard deviation bands) exclusion limits for Type-I (upper left), Type-II (upper right), flipped (lower left), lepton-specific (lower right) models, as a function of \cosba and \tanb. Contours are derived from the projection on the 2HDM parameter space for the $\mA = 300\GeV$ signal hypothesis. The excluded region is represented by the shaded gray area. The regions of the parameter space where the natural width of the \A boson $\Gamma_\A$ is comparable to the experimental resolution and thus the narrow width approximation is not valid are represented by the hatched gray areas.
  }
  \label{fig:cosba}
\end{figure*}

\begin{figure*}[!htb]\centering
    \includegraphics[width=0.495\textwidth]{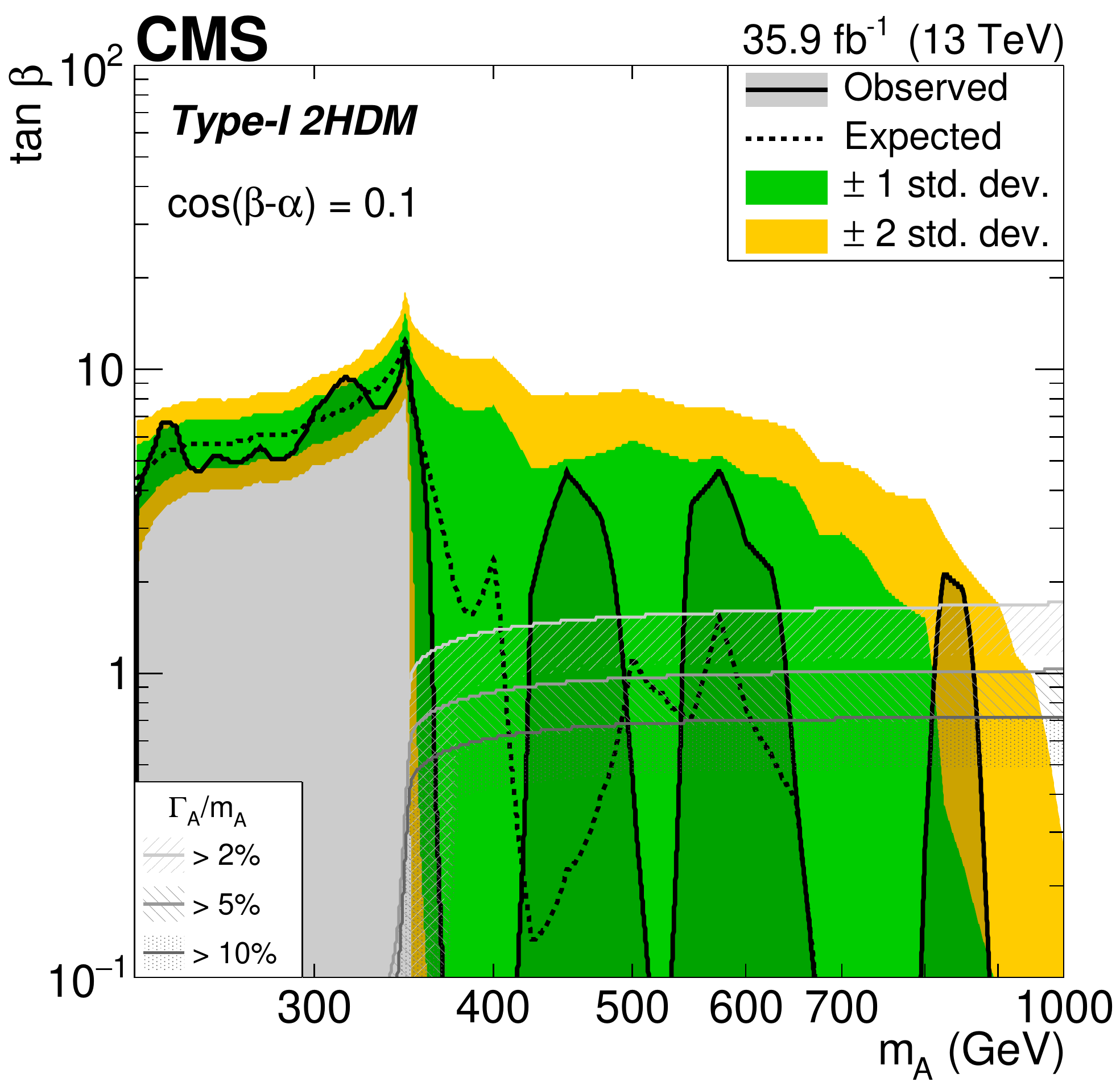}
    \includegraphics[width=0.495\textwidth]{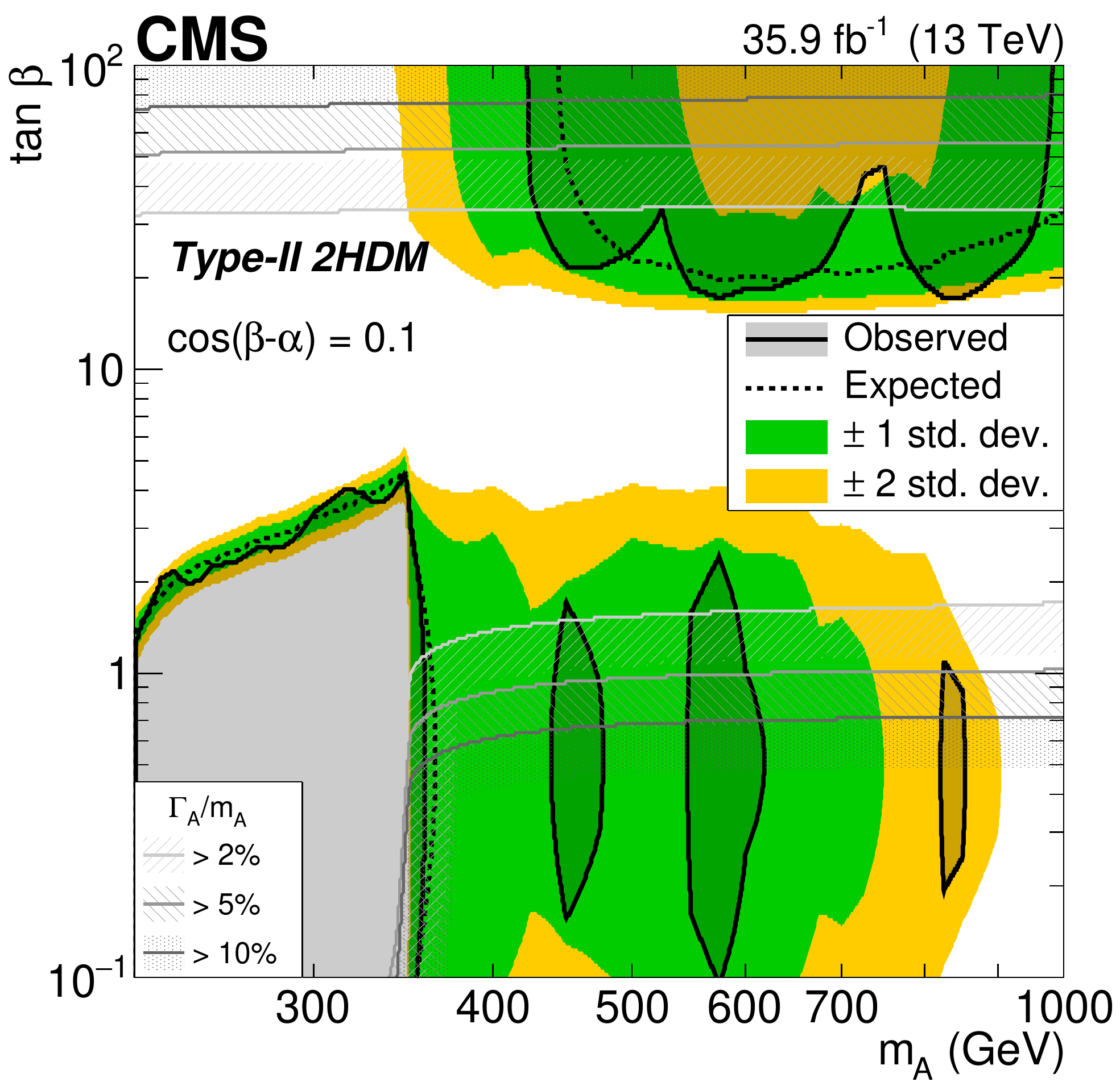}

    \includegraphics[width=0.495\textwidth]{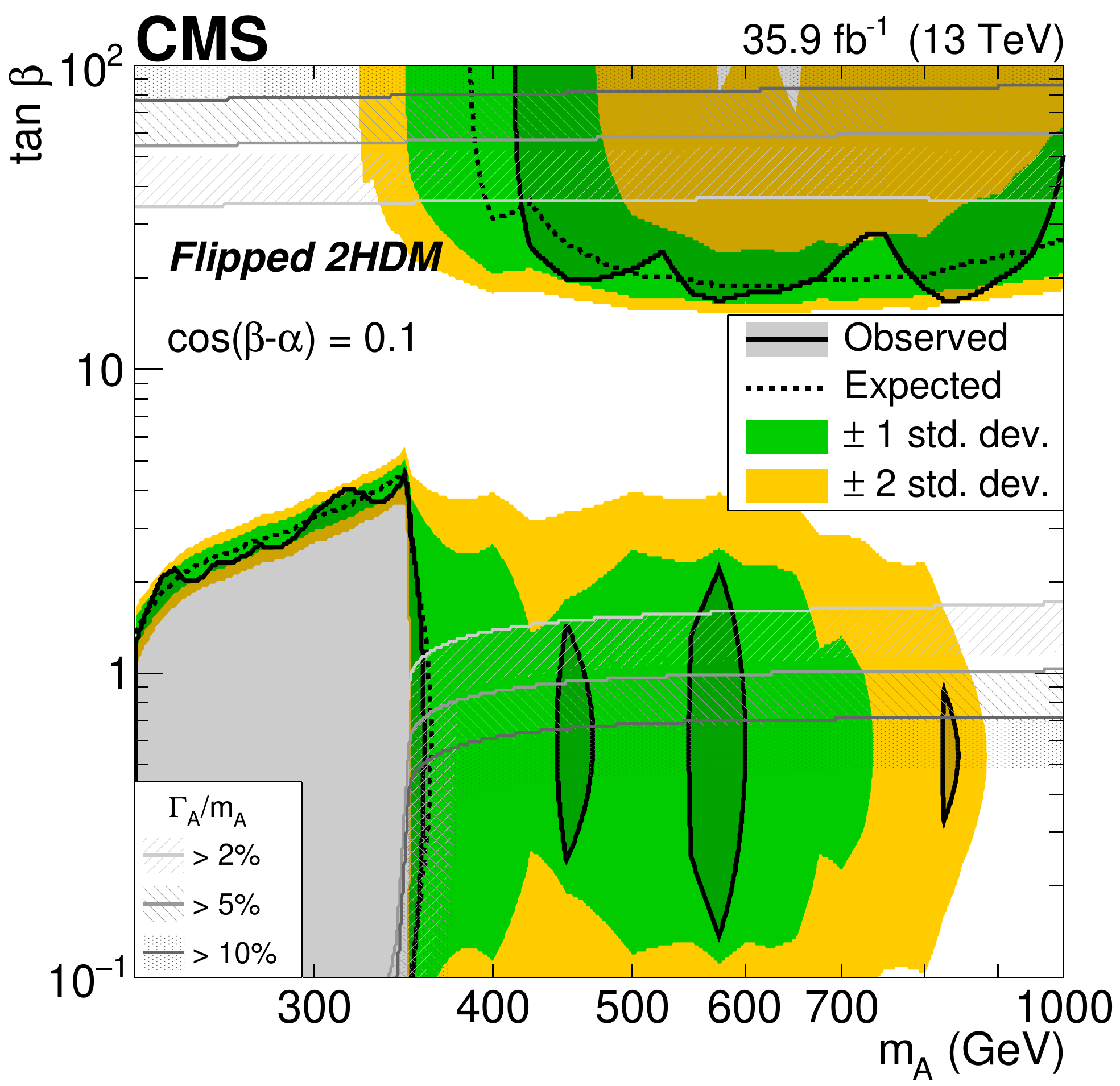}
    \includegraphics[width=0.495\textwidth]{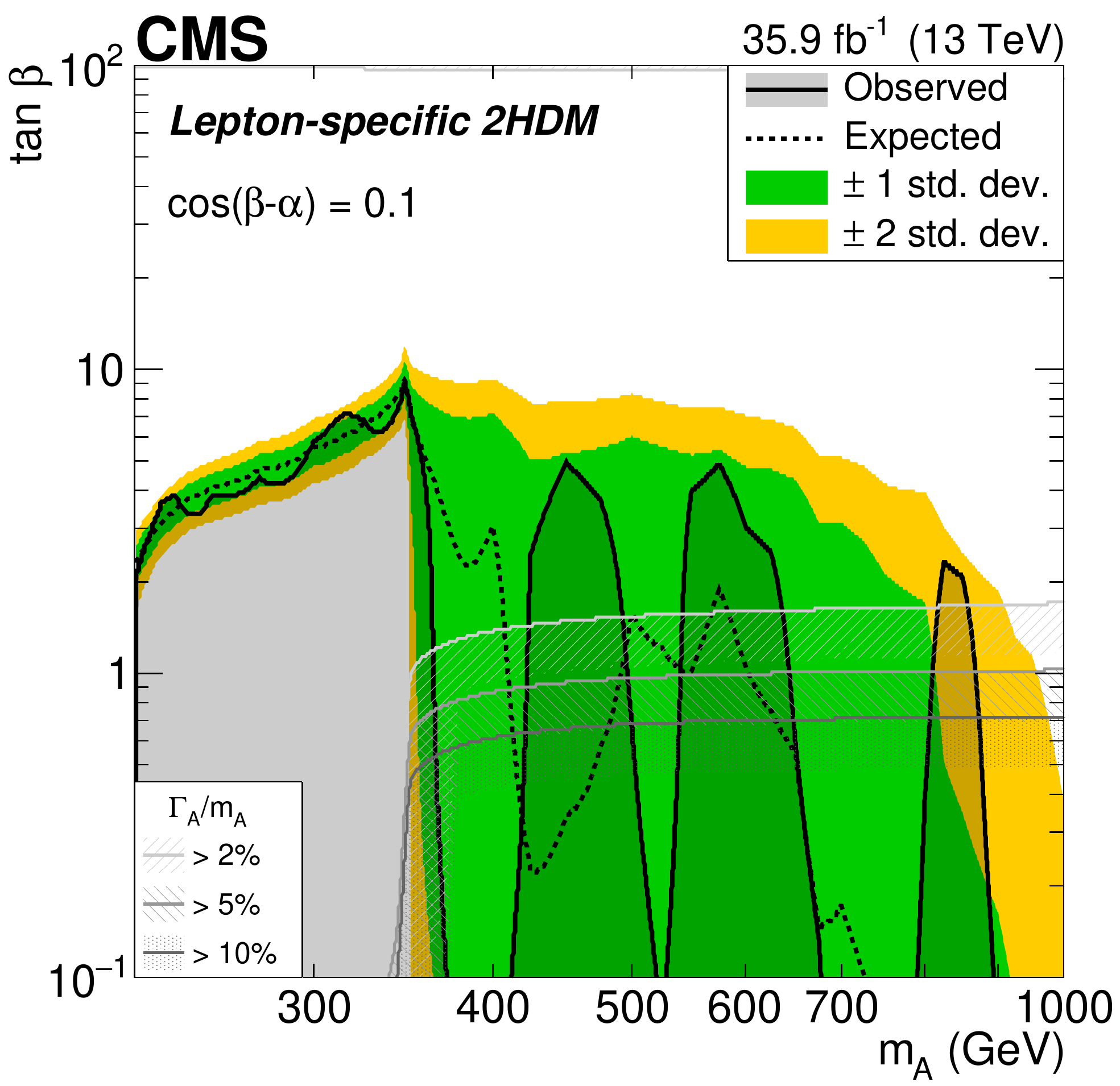}

   \caption{Observed and expected (with ${\pm}1,\,{\pm}2$ standard deviation bands) exclusion limits for Type-I (upper left), Type-II (upper right), flipped (lower left), lepton-specific (lower right) models, as a function of \mA and \tanb, fixing $\cosba = 0.1$. The excluded region is represented by the shaded gray area. The regions of the parameter space where the natural width of the \A boson $\Gamma_\A$ is comparable to the experimental resolution and thus the narrow width approximation is not valid are represented by the hatched gray areas.
  }
  \label{fig:mA}
\end{figure*}

The results are interpreted in terms of Type-I, Type-II, ``lepton-specific'', and ``flipped'' 2HDM formulations~\cite{Branco:2011iw}. 
In the scenario with $\cosba=0.1$ and $\tan\beta=3$, an \A boson up to $380$ and $350\GeV$ is excluded in 2HDM Type-I and Type-II, respectively, as depicted in Fig.~\ref{fig:limit}. These exclusion limits are used to constrain the two-dimensional plane of the 2HDM parameters $[\cosba, \tanb]$ as reported in Fig.~\ref{fig:cosba}, with fixed $m_\A=300\GeV$ in the range $0.1\le\tanb\le100$ and $-1\le\cosba\le1$, using the convention $0<\beta-\alpha<\pi$. Because of the suppressed \A boson cross section and $\B(\A\to\PZ\Ph)$, the region near $\cos(\beta-\alpha){\approx}0$ is not accessible in this search. On the other hand, $\B(\Ph\to\bbbar)$ vanishes in the diagonal regions corresponding to $\alpha$ close to $0$ in Type-II and flipped 2HDM, and $\alpha \to \pm \pi/2$ in Type-I and lepton-specific scenarios. The exclusion as a function of \mA, fixing $\cosba=0.1$, is also reported in Fig.~\ref{fig:mA}.

\section{Summary}\label{sec:conclusion}

A search is presented in the context of an extended Higgs boson sector for a heavy pseudoscalar boson \A that decays into a \PZ boson and an \Ph boson with mass of 125\GeV, with the \PZ boson decaying into electrons, muons, or neutrinos, and the \Ph boson into \bbbar. The SM backgrounds are suppressed by using the characteristics of the considered signal, namely the production and decay angles of the \A, \PZ, and \Ph bosons, and by improving the \A mass resolution through a kinematic constraint on the reconstructed invariant mass of the \Ph boson candidate.
No excess of data over the background prediction is observed. Upper limits are set at 95\% confidence level on the product of the \A boson cross sections and the branching fractions $\sigma_\A \,\B(\A\to\PZ\Ph) \,\B(\Ph\to\bbbar)$, which exclude 1 to 0.01~\unit{pb} in the 225--1000\GeV mass range, and are comparable to the corresponding ATLAS search.
Interpretations are given in the context of Type-I, Type-II, flipped, and lepton-specific two-Higgs-doublet model formulations, thereby reducing the allowed parameter space for extensions of the SM with respect to previous CMS searches.

\begin{acknowledgments}
We congratulate our colleagues in the CERN accelerator departments for the excellent performance of the LHC and thank the technical and administrative staffs at CERN and at other CMS institutes for their contributions to the success of the CMS effort. In addition, we gratefully acknowledge the computing centers and personnel of the Worldwide LHC Computing Grid for delivering so effectively the computing infrastructure essential to our analyses. Finally, we acknowledge the enduring support for the construction and operation of the LHC and the CMS detector provided by the following funding agencies: BMWFW and FWF (Austria); FNRS and FWO (Belgium); CNPq, CAPES, FAPERJ, and FAPESP (Brazil); MES (Bulgaria); CERN; CAS, MoST, and NSFC (China); COLCIENCIAS (Colombia); MSES and CSF (Croatia); RPF (Cyprus); SENESCYT (Ecuador); MoER, ERC IUT, and ERDF (Estonia); Academy of Finland, MEC, and HIP (Finland); CEA and CNRS/IN2P3 (France); BMBF, DFG, and HGF (Germany); GSRT (Greece); NKFIA (Hungary); DAE and DST (India); IPM (Iran); SFI (Ireland); INFN (Italy); MSIP and NRF (Republic of Korea); LAS (Lithuania); MOE and UM (Malaysia); BUAP, CINVESTAV, CONACYT, LNS, SEP, and UASLP-FAI (Mexico); MBIE (New Zealand); PAEC (Pakistan); MSHE and NSC (Poland); FCT (Portugal); JINR (Dubna); MON, RosAtom, RAS and RFBR (Russia); MESTD (Serbia); SEIDI, CPAN, PCTI and FEDER (Spain); Swiss Funding Agencies (Switzerland); MST (Taipei); ThEPCenter, IPST, STAR, and NSTDA (Thailand); TUBITAK and TAEK (Turkey); NASU and SFFR (Ukraine); STFC (United Kingdom); DOE and NSF (USA).

 \hyphenation{Rachada-pisek} Individuals have received support from the Marie-Curie program and the European Research Council and Horizon 2020 Grant, contract No. 675440 (European Union); the Leventis Foundation; the A.P.\ Sloan Foundation; the Alexander von Humboldt Foundation; the Belgian Federal Science Policy Office; the Fonds pour la Formation \`a la Recherche dans l'Industrie et dans l'Agriculture (FRIA-Belgium); the Agentschap voor Innovatie door Wetenschap en Technologie (IWT-Belgium); the F.R.S.-FNRS and FWO (Belgium) under the ``Excellence of Science -- EOS" -- be.h project n.\ 30820817; the Ministry of Education, Youth and Sports (MEYS) of the Czech Republic; the Lend\"ulet (``Momentum") Program and the J\'anos Bolyai Research Scholarship of the Hungarian Academy of Sciences, the New National Excellence Program \'UNKP, the NKFIA research grants 123842, 123959, 124845, 124850, and 125105 (Hungary); the Council of Science and Industrial Research, India; the HOMING PLUS program of the Foundation for Polish Science, cofinanced from European Union, Regional Development Fund, the Mobility Plus program of the Ministry of Science and Higher Education, the National Science Center (Poland), contracts Harmonia 2014/14/M/ST2/00428, Opus 2014/13/B/ST2/02543, 2014/15/B/ST2/03998, and 2015/19/B/ST2/02861, Sonata-bis 2012/07/E/ST2/01406; the National Priorities Research Program by Qatar National Research Fund; the Programa Estatal de Fomento de la Investigaci{\'o}n Cient{\'i}fica y T{\'e}cnica de Excelencia Mar\'{\i}a de Maeztu, grant MDM-2015-0509 and the Programa Severo Ochoa del Principado de Asturias; the Thalis and Aristeia programs cofinanced by EU-ESF and the Greek NSRF; the Rachadapisek Sompot Fund for Postdoctoral Fellowship, Chulalongkorn University and the Chulalongkorn Academic into Its 2nd Century Project Advancement Project (Thailand); the Welch Foundation, contract C-1845; and the Weston Havens Foundation (USA).

\end{acknowledgments}

\bibliography{auto_generated}
\cleardoublepage \appendix\section{The CMS Collaboration \label{app:collab}}\begin{sloppypar}\hyphenpenalty=5000\widowpenalty=500\clubpenalty=5000\input{HIG-18-005-authorlist.tex}\end{sloppypar}
\end{document}

%% file: HIG-18-005-authorlist.tex
\vskip\cmsinstskip
\textbf{Yerevan Physics Institute, Yerevan, Armenia}\\*[0pt]
A.M.~Sirunyan, A.~Tumasyan
\vskip\cmsinstskip
\textbf{Institut f\"{u}r Hochenergiephysik, Wien, Austria}\\*[0pt]
W.~Adam, F.~Ambrogi, E.~Asilar, T.~Bergauer, J.~Brandstetter, M.~Dragicevic, J.~Er\"{o}, A.~Escalante~Del~Valle, M.~Flechl, R.~Fr\"{u}hwirth\cmsAuthorMark{1}, V.M.~Ghete, J.~Hrubec, M.~Jeitler\cmsAuthorMark{1}, N.~Krammer, I.~Kr\"{a}tschmer, D.~Liko, T.~Madlener, I.~Mikulec, N.~Rad, H.~Rohringer, J.~Schieck\cmsAuthorMark{1}, R.~Sch\"{o}fbeck, M.~Spanring, D.~Spitzbart, A.~Taurok, W.~Waltenberger, J.~Wittmann, C.-E.~Wulz\cmsAuthorMark{1}, M.~Zarucki
% ...
\vskip\cmsinstskip
% ...
1:  Also at Vienna University of Technology, Vienna, Austria\\
% ...
